\documentclass{aa}

\usepackage{hyperref, makecell, soul}
\hypersetup{linkcolor=cyan,urlcolor=magenta}

\usepackage{amsmath}
\usepackage{enumitem}
\usepackage{graphicx}
\usepackage{multicol}
\usepackage{xcolor}

\newcommand\ciins{[CII]}
\newcommand\cii{\ciins\;}
\newcommand\ciiumns{\ciins-158$\mu$m}
\newcommand\ciium{\ciiumns\;}
\newcommand\Inuns{$\langle I_{\nu, \text{\ciins}}\rangle$}
\newcommand\Inu{\Inuns\;}
\newcommand\Lnuns{$L_{\nu, \text{\ciins}}$}
\newcommand\Lnu{\Lnuns\;}
\newcommand\simsthreens{\texttt{SIMSTACK3}}
\newcommand\simsthree{\simsthreens\;}

\newcommand{\linsimstackns}{\texttt{LinSimStack}}
\newcommand{\linsimstack}{\linsimstackns\;}

\newcommand{\reviewEditOne}[1]{#1}
\newcommand{\reviewEditTwo}[1]{#1}

\begin{document}

\title{Far-infrared lines hidden in archival deep multi-wavelength surveys: Limits on \ciium at $z\sim 0.3-2.9$}

\author{Shubh Agrawal\inst{1}\fnmsep\thanks{\email{shubh@sas.upenn.edu}; John Templeton TEX Fellow; Quad Fellow}
\and James Aguirre \inst{1} \and Ryan Keenan \inst{2}
}

\institute{Department of Physics and Astronomy, University of Pennsylvania, Philadelphia, PA 19104, USA
\and
Max-Planck-Institut für Astronomie, Königstuhl 17, D-69117 Heidelberg, Germany}

\date{Received July 19, 2025; Revised September 28, 2025, October 31, 2025; Accepted November 2, 2025; Published XXX}

\abstract{
\textit{Context:} Singly ionized carbon \cii is theorized to be the brightest emission line feature in star-forming galaxies, and hence an excellent tracer of the evolution of cosmic star formation. Archival maps from far-infrared and submillimeter surveys potentially contain the redshifted \ciiumns, hidden in the much-brighter continuum emission.

\textit{Aim:} We present a search for aggregate \ciium line emission across the predicted peak of star formation history by tomographically stacking a high-completeness galaxy catalog on broadband deep maps of the COSMOS field and constraining residual excess emission after subtracting the continuum spectral energy distribution (SED).

\textit{Methods:} The COSMOS equatorial $2\deg^2$ patch has been mapped by \textit{Spitzer}, \textit{Herschel}, and SCUBA2/JCMT. Using the high precision UV-O-IR photometry catalog COSMOS2020, we performed unbiased simultaneous stacking of $\sim360,000$ photometric redshifts on these confusion-limited maps to resolve the sub-THz radiation background. 
By subtracting a continuum SED model with conservative uncertainty estimation and completeness correction through comparison to the \textit{COBE}/FIRAS monopole spectrum, we obtain tomographic constraints on the sky-averaged \ciium signal within the three SPIRE maps: $11.8 \pm 10.2$, $11.0 \pm 8.7$, $9.6 \pm 9.8$, and $9.2 \pm 6.6$ $k$Jy/sr at redshifts $z\sim 0.65$, $\sim1.3$, $\sim2.1$, and $\sim2.6$, respectively, corresponding to $1-1.4\sigma$ significance in each bin. 

\textit{Results:} 
Our $3\sigma$ upper limits are in tension with past $z\sim 2.6$ results from cross-correlating SDSS-BOSS quasars with high-frequency \textit{Planck} maps, and indicate a much less dramatic evolution ($\sim$$\times 7.5$) of the mean \cii intensity across the peak of star formation history than collisional excitation models or frameworks calibrated to the tentative \textit{Planck}xBOSS measurement. We discuss this tension, particularly in the context of in-development surveys (TIM, EXCLAIM) that will map this \cii at high redshift resolution. 

\textit{Conclusion:} Having demonstrated stacking in broadband deep surveys as a complementary methodology to next-generation spectrometers for line intensity mapping, these novel methods can be extended to upcoming galaxy surveys such as \textit{Euclid}, as well as to place upper limits on fainter atomic and molecular lines.
}

\titlerunning{Line intensities from deep multi-wavelength surveys}
\authorrunning{S. Agrawal et al.}

\maketitle

\section{Introduction \label{sec:intro}}

Constraining the cosmic history of star formation and the chemical evolution of the interstellar medium (ISM) in galaxies is an intriguing and outstanding question in observational cosmology, forming the basis for in-development and planned surveys \citep{vieira2020tim, Pullen_2023, CCAT_Prime_Collaboration_2022,Keating_2015,Keating_2020}. Observations from the last decade suggest a consistent formalism in which the star formation rate (SFR) peaked roughly 3.5 Gyr after the Big Bang (the so-called ``cosmic noon''), at $z\sim2$ \citep{Madau_2014} and has decayed exponentially to the present day due to the decrease in available cold gas, as well as stellar winds and active galactic nuclei (AGNs). Dust in the ISM absorbs ultraviolet (UV) radiation from star-forming galaxies (SFGs), re-emitting it in the thermal far-infrared (FIR); half of the energy produced as starlight has been absorbed and re-emitted by dust as the cosmic infrared background (CIB) \citep{Dole_2006}.

Carbon is one of the most abundant metals in the ISM and singly ionizes to \cii (or C+) at 11.6~eV, a lower excitation energy than hydrogen. The \ciium fine structure line is empirically expected to be the primary coolant and brightest feature in SFGs, emitting up to 0.5-1\% of the total FIR luminosity \citep{1985crawford,1991ApJStacey}. Charting the redshift evolution of the mean \cii luminosity is an important diagnostic of the star formation rate density (SFRD) \citep{2021Yang, 2015Vallini, 2024Liang, Olsen_2017}. As a result, in addition to probing the growth of large-scale structure \citep{2022Karkare, 2019ApJMoradinezhad, 2017Fonseca}, line intensity mapping with atomic lines, such as \ciiumns, aims to constrain the \reviewEditOne{formation and chemical evolution of these galaxies}. 

The mean \cii luminosity is currently constrained by local $z\sim0$ measurements of the \ciium luminosity function (LF) from \textit{Herschel}/PACS observations \citep{2017Hemmati}, as well as by $z\sim4-6$ measurements of \ciiumns-LF by the ALPINE-ALMA (ALMA Large Program to Investigate C+ at Early Times) survey \citep{Yan_2020, ALPINE2021}. 
There are only a few measurements across cosmic noon when the SFR and the ISM evolve dramatically, as the \cii line is redshifted to
wavelengths that are difficult to observe
from the ground and require sub-orbital or space platforms. 
An example of this is provided by \citet{Pullen_2018} and \citet{Yang_2019} (hereafter P18 and Y19), who report a tentative detection of the \cii emission at $z\sim2.6$ in \textit{Planck} CIB maps by tomographically cross-correlating with quasars and CMASS galaxies in the Sloan Digital Sky Survey’s (SDSS-III) Baryon Oscillations Spectroscopic Survey (BOSS) DR12 spectroscopic redshift catalog. 

Theoretical and semi-analytical models (SAMs) for \cii emission and its relation to SFR, metallicity, and gas depletion rate depend on redshift. 
While there is a large scatter due to unconstrained model parameters, collisional excitation models such as \citet{2012Gong, 2015Silva} tend to make higher predictions for \ciium luminosity \Lnu than models based on scaling relations that link \Lnu to the SFR, such as \citet{2016Serra, 2015Yue}, which are calibrated by local luminosity function or SFRD measurements. 
The Y19+P18 measurement of \Inu is bright, corresponding to approximately 20\% of the total CIB (as measured by FIRAS/\textit{COBE} \citep{Fixsen_1998}; see Figure \ref{fig:constraints}) and favors certain parameterizations of collisional models. Empirical models \citep{Padmanabhan_2019} also attempt to calibrate the \cii luminosity - halo mass function to this measurement and forecast the detection capabilities of current-generation \ciium observatories at cosmic noon and during reionization.

Such excess \ciium emission would also be present in FIR or submillimeter observations of the CIB, e.g., from \textit{Herschel} or \textit{Spitzer}. However, while \textit{Planck} constructed all-sky maps in bands up to 857 GHz by observing tens of thousands of square degrees, FIR space observations have been limited to deep observations of smaller patches of the sky. This motivates exploring correlations on smaller scales using methods such as ``stacking", where the correlated signal is amplified using an ancillary source catalog. Stacking has been used successfully to constrain the properties of CIB and FIR sources \citep{Dole_2006, 2009Devlin, 2012Bethermin, 2017Wilson, 2022Romano, 2024Dunne}. In highly confused images (such as those produced by smaller-aperture dishes observing in the FIR), simultaneous stacking \citep{2013Viero} of a catalog, binned into subgroups based on emitter properties, has been effective in overcoming inherent biases and has been employed to understand infrared (IR) emission as a function of galaxy properties, particularly redshift \citep{2013Viero, Viero_2015, 2018Sun, 2020Duivenvoorden, Viero_2022}. Our methods extend \citet{Viero_2022}, who used analogous COSMOS\footnote{Cosmic Evolution Survey} field data to explore aggregate dust and thermal properties; we further analyze the residuals from a model fit to the continuum of the spectral energy distribution (SED), rather than the regressed model parameters. The residuals after continuum removal constrain spectral line emission.

In this paper, we demonstrate the application of simultaneous stacking with mid- to far-IR and submillimeter maps and a deep photometric survey, to place limits on the cosmic line emission from \ciiumns. Section \ref{sec:data} describes the COSMOS deep field, the COSMOS2020 high-completeness redshift catalog, and maps from instruments onboard \textit{Spitzer}, \textit{Herschel}, and the JCMT (James Clerk Maxwell Telescope). Section \ref{sec:methods} outlines our methods: stacking in the confusion limit with \linsimstack, estimating errors in stacked fluxes, constraining \cii emission as a residual in continuum fitting, and correcting for catalog incompleteness. We report and discuss the results in Section \ref{sec:results}, and conclude with Section \ref{sec:conclusion}.

\section{Data used \label{sec:data}}

The COSMOS field \citep{Scoville_2007} is a $\sim$2 deg$^2$ field in the equatorial sky, centered at RA $+150.12~\deg$, Dec $+2.21~\deg$ (J2000), which has been observed at accessible wavelengths from X-ray to radio by space- and ground-based observatories. For our analysis, the relevant data fall into the following two classes of products.

\begin{figure}[ht!]
\centering
\includegraphics[width=\linewidth]{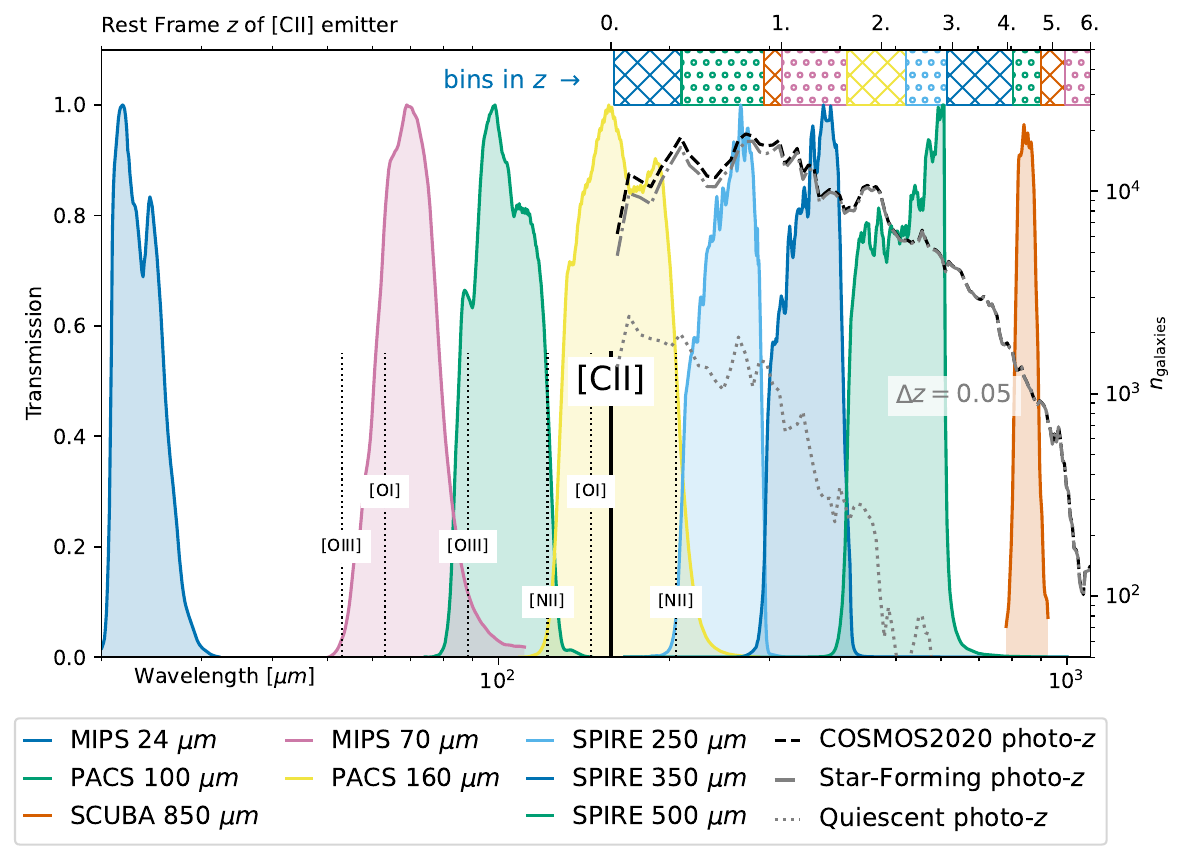}
\caption{\label{fig:filter_curves} Overview of data used in our analysis. We include transmission curves (left vertical, bottom horizontal axes) for the eight broadband maps used, from \textit{Spitzer}, \textit{Herschel}, and SCUBA-2. We overplot the rest-frame wavelengths of selected FIR emission lines, including \ciiumns; these lines are redshifted into the broadband maps. The top horizontal axis labels translate the observer wavelengths into \reviewEditOne{the} rest-frame redshift $z$ of a \ciium emitter. We trace the $z$ distribution of the COSMOS2020 photometric catalog (in bins of $\Delta z = 0.05$), with number counts on the right vertical axis; \reviewEditOne{number counts for the selection of star-forming or quiescent emitters are also shown}. Redshifted C+ emission will appear in the SPIRE maps for specific $z$ ranges. Finally, at the top, we demarcate the $z$-binning used for stacking with \linsimstack; bins at $z\sim0.65$, 1.3, 2.1, and 2.6 were chosen to overlap with the SPIRE bands.}
\end{figure}

\textbf{Maps:} For our search for a line-emission signal, we required maps such that \texttt{(a)} a model for the continuum emission could be constrained in our wavelength regime and \texttt{(b)} the \ciium signal from a subset of our galaxy catalog would potentially be redshifted into a map's broadband filter. Table \ref{tab:maps} lists the eight maps from mid-IR to submillimeter that we used to bracket the emission SED. The maps were obtained from the \reviewEditTwo{SEIP\footnote{Spitzer Enhanced Imaging Products} \citep{ssc_and_irsa_spitzer_2020}, HELP\footnote{Herschel Extragalactic Legacy Project} \citep{2021MNRAS.507..129S}, and SCUBA-2 \citep{2017MNRAS.465.1789G}} \footnote{Submillimetre Common-User Bolometer Array 2; \href{https://zenodo.org/records/57792}{zenodo.org/records/57792}} archives. If necessary, we applied color corrections and divided the map by the beam obtained from auxiliary data products. 

\begin{table}[htb]
\caption{Eight broadband maps of the COSMOS field used in the analysis.}
\begin{tabular}{c|c|c|c}
\centering
Instrument & $\lambda$ ($\mu$m) & FWHM & Color Corr.%& r.m.s. ($m$Jy) 
\\ \hline
\textit{Spitzer}/MIPS & 24 & 5.51" & 1.24 %& .0279 
\\
\textit{Spitzer}/MIPS & 70 & 18.18" & 1.32 %&  
\\
\textit{Herschel}/PACS & 100 & 7.49" & - %& 0.10 
\\
\textit{Herschel}/PACS & 160 & 11.33" & - %& 2.11 
\\
\textit{Herschel}/SPIRE & 250 & 17.62" & - %& 6.89 
\\
\textit{Herschel}/SPIRE & 350 & 24.42" & 0.9914 %& 7.40 
\\
\textit{Herschel}/SPIRE & 500 & 35.69" & 0.95615 %& 7.39 
\\
JCMT/SCUBA-2 & 850 & 12.1" & - %& 8.07 
\end{tabular}
\\
Notes: {We list the nominal band wavelength, the beam size (FWHM) of the instrument in the band (measured from auxiliary beam calibration data products, as in \citet{Viero_2022}), and the color correction factors applied \citep{2010A&A...512A..78B, Viero_2022}.}
\label{tab:maps}
\end{table}

\textbf{Photometry Catalog.} 
The top and right axes of Figure \ref{fig:filter_curves} trace the number density of emitters (dash-dotted black line) in the COSMOS2020 catalog as a function of redshift. Histograms of \reviewEditOne{the star-forming and quiescent selection of galaxies are also shown individually.} Notably, \ciium emission can be measured in all three SPIRE maps \footnote{\reviewEditOne{ Figure \ref{fig:filter_curves} also shows that narrow spectral features other than \ciium could be redshifted into the SPIRE bands (e.g., [OI] and [NII] lines), which correlate with the emitters within a redshift bin and could potentially contaminate our measurement. This is further discussed in Section \ref{subsec:contamination}.}}. Similar to \citet{Viero_2022}, we used the FARMER photometric redshifts, which are typically expected to be more accurate at our redshifts and for fainter sources.

\section{Methods \label{sec:methods}}
\subsection{Stacking in the confusion limit with \linsimstack \label{subsec:simstack}}

Far-IR or submillimeter observations from single-dish telescopes, such as the maps discussed in Section \ref{sec:data}, are restricted in angular resolution, resulting in the so-called ``confusion limit". Simultaneous stacking has been shown to overcome the clustering bias inherent in stacking on maps dominated by confusion noise \citep{2013Viero, Viero_2022}, unlike conventional postage-stamp stacking (mean or median) that is prone to wavelength-dependent biases. 

We modeled a broadband map $m_\lambda$ (in Jy/beam, indexed by the mean band wavelength $\lambda$) as a linear combination of several sub-maps or layers, each layer $A^{z, m}$ being a convolution of the instrument beam with a hits-map of a portion or bin of the galaxy catalog. The binning was user-defined and based on properties that determine emission in that band, such that stacking with these homogeneous sources gives a meaningful result. The coefficients $\bar{S}_{z, m}(\lambda)$, thus regressed, give the average flux density of emitters in that bin or layer:
\begin{align}
\label{eqn:linear}
m_{\lambda} &= A^{z, m} \bar{S}_{z, m}(\lambda) + \epsilon \\
\Rightarrow \bar{S}_{z, m}(\lambda) &= (A^T W A)^{-1} A^T W m_{\lambda},
\end{align}
where $W$ is the weight matrix, given by the inverse per-pixel variance in the broadband maps, and $\epsilon$ is the corresponding Gaussian noise vector. The linearity of the regressed parameters $\bar{S}_{z, m}(\lambda)$ can be exploited to write down a closed form for the least-squares minimization problem.

For our application, we developed \linsimstack \footnote{\href{https://github.com/shubhagrawal30/LinSimStack}{github.com/shubhagrawal30/LinSimStack}} as a fork of the previously introduced Python-based \simsthree \citep{Viero_2022}. Our version addresses inefficiencies in the \simsthree codebase, notably exploiting this linearity by analytically marginalizing over the coefficients, instead of using a Levenberg-Marquardt non-linear solver, resulting in order-of-magnitude gains in speed with \linsimstackns  \footnote{Other improvements included code profiling and refactoring to improve time and memory complexity bottlenecks; discussion of these is outside the scope of this work and further implementation details are available on GitHub.}. This also enables us to analytically estimate the statistical covariance (see Section \ref{subsec:errors}).

We also implemented the following additions, some of which have been utilized in previous work. We included an additive flat foreground layer in Eqn. \ref{eqn:linear} to mitigate biases from mean subtraction of the maps or galactic foreground contamination. We additionally followed \citet{2020Duivenvoorden} to model the leakage of flux from masked areas due to large beams; we added another layer that is the beam convolved with the masked pixels of the galaxy catalog. All redshift and stellar mass bins were simultaneously stacked to reduce any bias due to interlopers \citep{2018Sun}. Finally, linear least-squares fitting was performed twice, with bright outlier pixels removed in the second iteration to reduce positive bias from close interlopers or bad pixels. 

\begin{figure}[ht!]
\centering
\includegraphics[width=\linewidth]{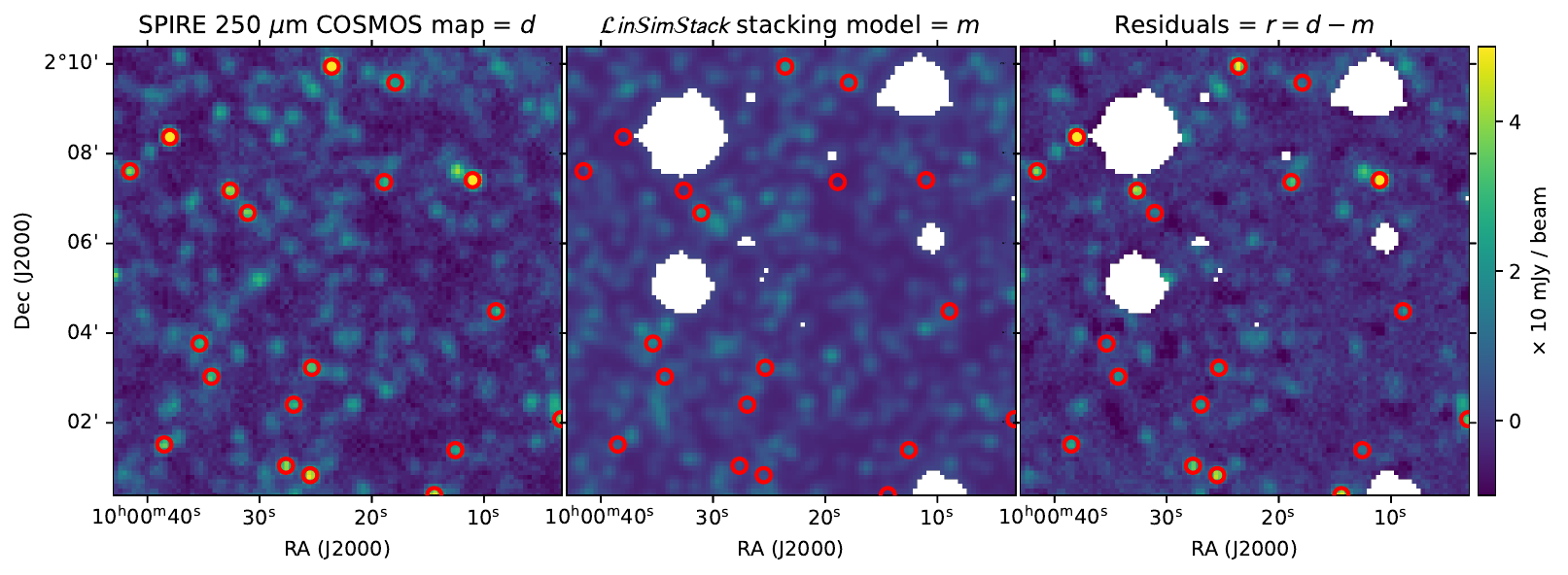}
\caption{\label{fig:stamps} Zoom-in on the stacking outputs for the \textit{Herschel}/SPIRE 250 $\mu$m COSMOS map, illustrating our forward modeling \linsimstack methodology. Left to right: 2D vectors for the data $d$, model $m$, and the residuals $r=d-m$. The model $m$ is a linear combination of all beam-convolved hit-maps constructed using the COSMOS2020 photometric catalog. We also plot sources from the \textit{Herschel}/SPIRE Point Source Catalog (HSPSC) with estimated fluxes in the HSPSC above 30 mJy, which correlate with the residual flux in the 2D vector $r$ (see Section \ref{sec:completeness}, Appendix \ref{app:linsimstack}, the corresponding figures, and the discussion for further details on model incompleteness).}
\end{figure}

We binned our galaxy catalog based on three observables, which are expected to result in similar spectral energy distributions (SEDs) in the thermal IR \citep{2018Schreiber}. We divided the sample into star-forming and quiescent emitters using the rest-frame two-color $NUV-r$ versus $R-J$ criterion, formulated in \citet{Ilbert2013, 2013Arnouts} and used by \citet{weaver_cosmos2020_2022, Viero_2022}. We binned by redshift \reviewEditOne{$z=0.01-6$}, both because the SED is redshifted and galaxy rest-frame emission changes with cosmic evolution; relevant bin edges in $z$ were selected to match the full width at half maximum (FWHM) of the \textit{Herschel}/SPIRE transmission (see top of \ref{fig:filter_curves}).
Finally, we binned the catalog by the (logarithm of) stellar mass $\log(M/M_\odot)$, as it is a good tracer of IR properties \citep{2013Viero, Viero_2022}. Although \citet{Viero_2022} only works with $\log(M/M_\odot) = 9.5-12$, we binned across a larger range, $8.5-13$, as we aim for the catalog to describe as much FIR emission as possible with COSMOS2020. 
\footnote{Note that \citet{2020Duivenvoorden} were able to resolve a large fraction of the CIB with sources that are part of the COSMOS2020 catalog; unlike their analysis, we do have to exclude a portion that does not have estimates for photometric redshifts.} 
Our bin definitions and the corresponding number counts per bin are detailed in Tables \ref{table:N_star} and \ref{table:N_nonstar} for the star-forming and quiescent samples, respectively. We empirically find that increasing the total number of sources by including higher redshifts or lower stellar masses does not improve completeness metrics (fraction of the CIB resolved; Section \ref{sec:completeness})

\subsection{Unbiased error estimation \label{subsec:errors}}
We estimated errors in the average emission flux, $\bar{S}_{z, m}(\lambda)$, measured by \linsimstack using the following four independent methods, and conservatively combined them to obtain a final variance in $\bar{S}$. 

\paragraph{Linear formulation:} For the linear least-squares problem, the statistical covariance has the analytical form $C_{\bar{S}} = (A^T W A)^{-1}$.
This provides a robust lower bound for the complete error in $\bar{S}_{z, m}(\lambda)$, as it reflects the random error limit resulting from the covariance of the map pixels, but does not include any information about systematics. 
\paragraph{Pixel bootstrapping:} Similarly to \citet{2020Duivenvoorden}, we randomly selected $80\%$ of the unmasked pixels and performed the least-squares stacking algorithm ten times to randomly resample the pixels in each map, corresponding to different instrumental noise realizations. 
\paragraph{Catalog bootstrapping:} We followed \citep{Viero_2022} and ran ten iterations with $80\%$:$20\%$ split bootstrapping on the catalog, doubling the number of bins while still fitting all emitters available to prevent negatively biasing the fluxes. Galaxy bootstrapping also captures the noise-realization variation encoded by pixel bootstrapping; the variance in $\bar{S}_{z, m}(\lambda)$ when randomly sampling over galaxies is expected to be higher than when pixel sampling.
\paragraph{Jack-knifing:} Similarly to \citet{2020Duivenvoorden}, for a given map, we reran our stacking algorithm four times, with each quartile of the map masked for one run. This encapsulates sample, or cosmic, variance at the scale of the field ($\lesssim$ 2 sq. deg.). 

\begin{figure}
    \centering
    \includegraphics[width=\linewidth]{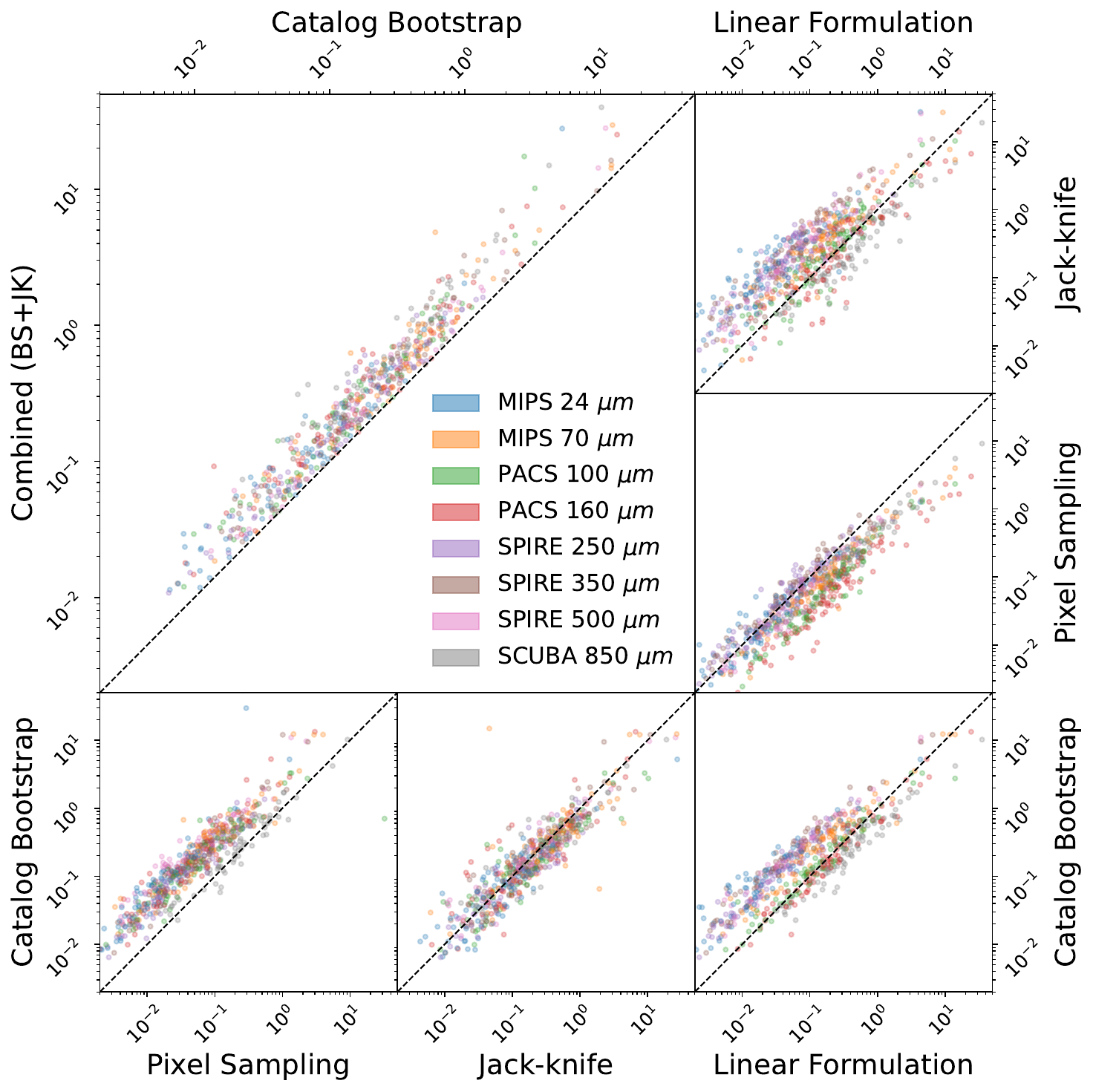}
    \caption{Errors in stacked mean intensities estimated via four methods (see Section \ref{subsec:errors}). Fractional errors in the stacked fluxes, obtained as outputs of the \linsimstackns are color-coded by band. The statistical lower bound is given by the closed-form error formulation, obtained by writing the stacking problem as a linear regression. Bootstrapping the catalog or the pixels captures the additional systematic errors, as it samples different realizations of pixel noise and emitter variance; empirically, the bootstrapping over the catalog (lower left subplot) typically yields the more conservative error estimates. Jack-knifing (JK) provides an estimate of the cosmic variance (CV) at the scale of the map. We conservatively add in, in quadrature, the JK estimate with the maximum of the other three estimates. This effectively lower-bounds our bootstrapping error estimate by the analytical linear formulation limit. Past analyses \citep{Viero_2022, 2013Viero} have only used catalog bootstrapping as the error estimation method. The main subplot above (top-left) compares our final error estimates with those from catalog bootstrapping alone. Our estimates are more conservative by definition and additionally include a measure of CV.}
    \label{fig:errors}
\end{figure}

Figure \ref{fig:errors} shows the fractional error estimates in the mean stacked fluxes using these four individual methods. The uncertainty in $\bar{S}_{z, m}(\lambda)$ is one of two direct contributors to the constraining power on \cii emission, the other being the uncertainty in the SED fit (see Sections \ref{subsec:continuum} and \ref{sec:cii}). Empirically, the overall uncertainty in our line emission measurement is moderately more dependent on the stacking flux uncertainty than the SED fit uncertainty. 
We conservatively selected the quadrature sum of the jackknifing error (which encodes cosmic variance on the field scale) and the maximum of the other three error estimates, to obtain $\sigma(\bar{S}_{z, m}(\lambda))$ for subsequent analyses.

\subsection{Fitting SED continuum models \label{subsec:continuum}}

\begin{figure*}
\centering
\includegraphics[width=\linewidth]{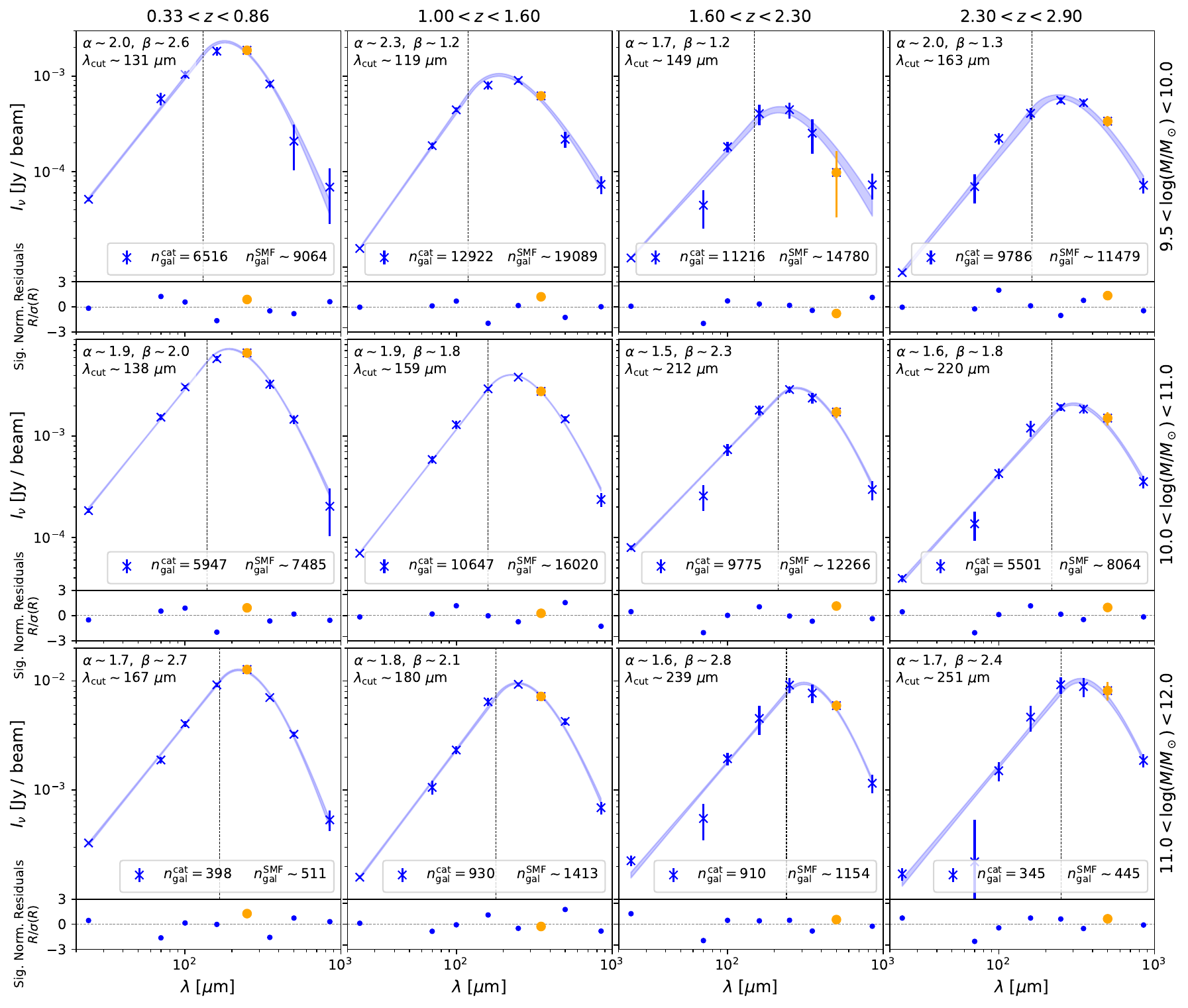}
\includegraphics[width=\linewidth]{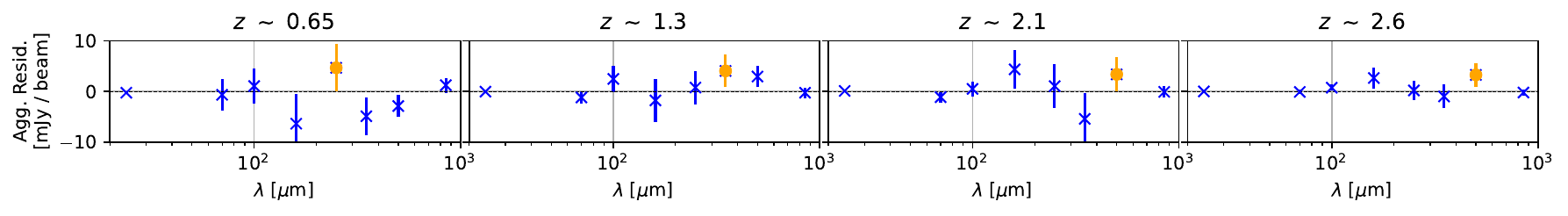}
\caption{\label{fig:sed_fits}Stacked intensity estimates obtained from \linsimstack for individual  COSMOS2020 bins, fit with a modified black-body emission continuum SED model. The top panels of each sub-figure show the stacked mean intensities (blue crosses) and corresponding $1\sigma$ envelope of the \texttt{emcee} fits (blue) for a subset of the catalog bins; the bottom panels show the signal-to-noise of the fit residuals. The vertical dashed line represents the wavelength cut, $\lambda_\text{cut}$, between the two piecewise components of the SED continuum model in Eqn. \ref{eqn:sed} (i.e., the frequency $\nu_0$). For a specific $z$ interval, \ciium emission is redshifted into a SPIRE band and manifests as excess residual emission over the continuum; these are marked in orange. 
\reviewEditOne{The SPIRE 500$\mu$m map contains redshifted \ciium from the highest two $z$ bins.}
Different rows and columns correspond to different stellar mass $\log(M/M_\odot)$ and redshift $z$ cuts; all bins shown here are from the star-forming selection. The labels indicate the number count within a bin in the catalog and as predicted by the stellar mass function (SMF) from \citet{Weaver_2023}. We conservatively inflate the variance on the \ciium residual emission by the reduced chi-squared, $\chi_r^2$, statistic of each SED fit. The bottom row shows residuals obtained by adding over all bins at the same redshift interval, with potential \ciium emission again marked in orange. \\
\reviewEditOne{For brevity, we only show stacking within the three highest stellar mass bins (for each redshift bin with potential \ciium emission); these are the predominant contributors to the CIB and \ciium emission. (See Appendix \ref{app:allstacking} for stacking results in all bins.) The bottom row shows the residuals obtained by adding all stellar mass and star-forming plus quiescent bins} at a redshift.} 
\end{figure*}

We obtained stacked flux densities, $\bar{S}_{z, m}(\lambda)$,  from Section \ref{subsec:simstack} and a diagonal covariance matrix from Section \ref{subsec:errors}. The values of $\bar{S}_{z, m}(\lambda)$ trace the SED of emitters in each grouped sub-catalog, which is a superposition of several black bodies. This is well modeled \citep{2013Viero, Viero_2022, Pullen_2018} by a modified black-body function,

\begin{equation}
\label{eqn:sed}
S_{\nu = c/\lambda}(\Theta) \propto \begin{cases}
\nu^\beta B_\nu (T); & \nu < \nu_0 \\
\nu^{-\alpha}; & \nu > \nu_0 \\
\end{cases},
\end{equation}
\\
where $B_\nu$ is the nominal Planck function, $T$ is the average dust temperature in the bin, $\nu_0$ is the transition between the mid-IR Wien and the Rayleigh-Jeans side. The exponent of the Wien side is given by  $\alpha$, while $\beta$ is the emissivity index. A single parameter, $A$, determines the overall amplitude of $S_\nu$, with the two functional pieces and their first derivatives required to be equal at $\nu_0$ (which removes one degree of freedom and sets $\nu_0$). The parameters of the SED model, $\Theta = (A, \alpha, \beta, T)$, were fit to the measured stacked fluxes using the Python MCMC (Markov chain Monte Carlo) package, \texttt{emcee} \citep{Foreman-Mackey_2013}, which samples the parameter space to compute the posterior for $\Theta$. We followed \citet{Viero_2022} for additional details while computing our Bayesian solution: 
\reviewEditOne{\texttt{(a)} we used their prescription for initial placement of walkers; \texttt{(b)} flux measurements consistent with zero were treated as upper limits, with their likelihood contribution considered via a survival function \citep{2012Sawicki}; \texttt{(c)} we inflated errors in the stacked MIPS 24~$\mu$m measurements, to account for polycyclic aromatic hydrocarbon (PAH) emission in massive galaxies ($\log M /M_\odot \geq 10$) by factors of 2, 4, and 2.4 for the $0.5<z<1.5$, $1.5<z<2$, and $2<z<2.5$ bins, respectively. \footnote{\reviewEditOne{We point out that PAH emission, mostly $3-20~\mu$m rest wavelengths, is expected to significantly contaminate only the MIPS 24$\mu$m map. Constraining the fitted temperature of the continuum SED was an integral goal of the \citet{Viero_2022} analysis, making biases in their lowest wavelength band crucial. However, given that the SPIRE bands lie predominantly on the Rayleigh-Jeans side of the SED, we do not expect uncertainty in the 24$\mu$m stacking measurement to significantly affect our residual measurement.}}. This is identical to \citet{Viero_2022}} \footnote{\href{https://github.com/marcoviero/simstack3/blob/37ef14b1a28093c81ff2a7e540a48cc073104b8d/simstackcosmologyestimators.py\#L342}{SIMSTACK implementation, open-sourced on GitHub}}

We adopted wide uniform priors for all parameter: $\log(A) \in \mathcal{U}(-10, 0)$, $\alpha \in \mathcal{U}(1, 3)$, $\beta \in \mathcal{U}(1, 3)$, and $T \in \mathcal{U}(5, 40)$. These priors are wide and encapsulate prior measurements for the physically meaningful $\alpha, \beta$, and $T$ \citep{2012MNRAS.425.3094C, 2014Planck}. For stability of the MCMC chains, we defined the overall normalization parameter, $A$, as the logarithm of the median of the modeled SED values at each of the eight central wavelengths.

\subsection{Aggregate IR background as a measure of completeness correction \label{sec:completeness}}
Within selected frequency bandwidths, we used \linsimstackns to measure the integrated average emission from galaxies within a redshift range $z$ and stellar mass $m$: $\bar{S}_{z, m}(\lambda)$. The portion of the CIB resolved by stacking with our catalog can be written as a sum over the bins,
\begin{equation}
F(\lambda) = \int dN_{z, m} \; \bar{S}_{z, m}(\lambda) \approx \sum_{z, m} N_{z, m} \bar{S}_{z, m}(\lambda).
\end{equation}
Here, $N_{z, m}$ is the number of emitters in a given bin. Naively, this can be set to the galaxy counts in COSMOS2020 input to \linsimstack, but this would be a significant underestimate due to incompleteness of the galaxy catalog. Instead, we used completeness-corrected measurements of the stellar mass function (SMF) $\frac{dN}{dm}(m, z)$, which encode the number density of emitters as a function of stellar mass content, at a given redshift:
\begin{align}
F(\lambda) = \int dV(z) n_{z, m} &\bar{S}_{z, m}(\lambda)
\\
= \iint dV(z)\; dm \frac{dN}{dm}(m, z) &\bar{S}_{z, m}(\lambda) 
\\
\label{eqn:SMF_corr1}
\approx \sum_z \sum_m \Big( \Delta m \frac{dN}{dm}(m, z) \Big) \Big( \delta V(z) \Delta z \Big) &\bar{S}_{z, m}(\lambda)
\\
\label{eqn:SMF_corr2}
= \sum_{z, m} N_{z, m}^\text{SMF-pred} &\bar{S}_{z, m}(\lambda).
\end{align}

Here, the terms, respectively, encode the expected number density of emitters in a given bin, the effective cosmological volume as an integral over the differential comoving volume, and the \linsimstack estimate for mean emission for this kind or bin of emitters. \citet{Weaver_2023} measures the SMF for star-forming and quiescent galaxies up to $z \lesssim 7.5$ using COSMOS2020, constraining the parameters of a double and single Schechter function, respectively. We adopted values for the maximum likelihood solution from Tables C.2 and C.3  of \citet{Weaver_2023} to compute the expected number of emitters in each bin. 

We report the predicted number counts from the SMF integration in brackets in Tables \ref{table:N_star} and \ref{table:N_nonstar} and select the higher of the two values (from our catalog and the SMF prediction) to prevent underestimation bias at higher redshifts.   
For the three Herschel/SPIRE maps, in which we report \cii emission, we resolve approximately $87\%$, $76\%$, and $73\%$ of the CIB, compared to \citet{Fixsen_1998}'s COBE/FIRAS measurements in the 250$\mu$m, 350$\mu$m, and 500$\mu$m bandpasses \footnote{These completeness fractions are lower than in the \citet{2020Duivenvoorden} analysis of the COSMOS field, because we only stack objects with valid photometric redshift solutions to enable tomography.}. 
We scaled the \cii limits we report (obtained using Eqn. \ref{eqn:cii}; see Section \ref{sec:cii}) by an inverse of this completeness fraction, assuming that a similar fraction of the missing cosmic background is \cii emission \footnote{We discuss this assumption further in Section \ref{sec:results} in the context of tension with prior measurements of \ciium at $z\sim 2.6$.}. 
Potential causes for this incompleteness are discussed further in Appendix \ref{app:linsimstack}, where we note that \linsimstack fails to resolve some of the brighter sources in the SPIRE maps, which correlate with sources present in the Herschel SPIRE Point Source Catalog (HSPSC). Together, the HSPSC sources in COSMOS account for $10-16\%$ of the FIRAS CIB monopole spectrum. 

\subsection{Aggregate \cii emission in Herschel/SPIRE maps \label{sec:cii}}
We fitted the parameters $\Theta$ of a modified black-body SED model, ${S}_{z, m}(\lambda; \Theta)$, which provides an estimate of the continuum emission in a given wavelength bin, $\lambda$.  Redshifted \reviewEditOne{\ciium} in a SPIRE map appears as residual emission over the fitted continuum
\begin{equation}
S^{\text{\cii}}_{z, m}(\lambda) = \bar{S}_{z, m}(\lambda) - {S}_{z, m}(\lambda; \Theta)
\label{eqn:residuals},
\end{equation}
where $\lambda$ and $z$ are related through the rest-frame emission frequency of \ciium, $\lambda_{\text{map}} = (1+z) 158\;\mu$m. The SPIRE maps thus contain \cii emission from $z \sim 0.5-3$, which spans the theorized peak of star formation history. 

The total intensity history of line emission is expressed in the same form as Eqns. \ref{eqn:SMF_corr1} and \ref{eqn:SMF_corr2}, with  normalization over the COSMOS field ($1.6 \deg^2$) area necessary to obtain the units of Jy/str:
\begin{align}
\label{eqn:cii}
I^{\text{\ciins}}_\nu(z_\text{\ciins}) = \sum_{z = z_\text{\ciins}, m} N_{z, m}^\text{SMF-pred} \times S^{\text{\cii}}_{z, m}(\lambda).
\end{align}
As noted in Section \ref{sec:completeness}, these values are scaled by the completeness fraction, i.e., the fraction of the CIB that our stacking was able to resolve. The uncertainty in the measurement are similar written as

\begin{align}
\label{eqn:sigcii}
\nonumber\sigma\Big(I^{\text{\ciins}}_\nu&(z_\text{\ciins})\Big) = \\
&\sum_{z = z_\text{\ciins}, m} N_{z, m}^\text{SMF-pred} \times \sigma\Big(S^{\text{\cii}}_{z, m}(\lambda)\Big)
\times \sqrt{\chi^2_{r; z, m}},
\end{align}

where $\sigma(S^{\text{\cii}}_{z, m}(\lambda))$ is the quadrature sum of uncertainties in the two terms of Eq. \ref{eqn:residuals} and $\chi^2_{z, m}$ is the reduced chi-squared statistic for each SED continuum model (excluding the data point that could contain redshifted \cii emission). We therefore conservatively scale our \cii uncertainties by a measure of the goodness of the fit of the model. We note empirically that, for certain galaxy bins, the SED models cannot be well constrained because there are insufficient detections of the mean emissions (i.e., the \linsimstack results are consistent with zero for several broadband maps). For these bins, we did not include a mean emission contribution in \ref{eqn:cii}, but we included an uncertainty contribution in \ref{eqn:sigcii} by setting $\sigma \Big(S^{\text{\cii}}_{z, m}(\lambda) \Big) = \sigma\Big(\bar{S}_{z, m}(\lambda)\Big)$. These bins contribute $\lesssim 0.06$\% of the FIRAS CIB monopole measurement (i.e., well within our quoted error bars), as they typically contain significantly fewer galaxies.

\section{Results and discussion \label{sec:results}}

\subsection{\reviewEditOne{\ciium constraints} \label{subsec:ciiconstraints}}

\begin{table}[]
\caption{Constraints on \ciium in four tomographic bins.}
    \begin{tabular}{c|c|c|c}
    \centering
         $z_{\text{median}}$ & $(z_{\text{min}}, z_{\text{max}})$ &  $I^{\text{\ciins}}_\nu$ ($k$Jy/sr) & $3\sigma$ upper bound
         \\\hline
         .65 & $(.33, .86)$ & $11.8 \pm 10.2$  & $<$ 41 $k$Jy/sr 
         \\
         1.3 & $(1.0, 1.6)$ & $11.0 \pm 8.7$ & $<$ 37 $k$Jy/sr
         \\
         2.1 & $(1.6, 2.3)$ & $9.6 \pm 9.8$ & $<$ 39 $k$Jy/sr
         \\
         2.6 & $(2.3, 2.9)$ & $9.2 \pm 6.6$ & $<$ 29 $k$Jy/sr
    \end{tabular}
    \\
    Notes: {The median redshift of emitters within the given redshift range in COSMOS2020 is $z_{\text{median}}$, while ($z_{\text{min}}$, $z_{\text{max}}$) are the bin edges used within the \linsimstack stacking framework. The uncertainties are quoted at $1\sigma$, and are propagated from uncertainties in the stacked flux estimates and the SED fit, weighted by number counts in each bin. The tomographic measurements range have S/Ns between $0.96-1.4$, with a total S/N of 2.42. Our constraints, particularly the $3\sigma$ upper limits, indicate a preference for the lower intensity models of \ciium intensity history.}
    \label{tab:constraints}
\end{table}

\begin{figure}
\centering
\includegraphics[width=\linewidth]{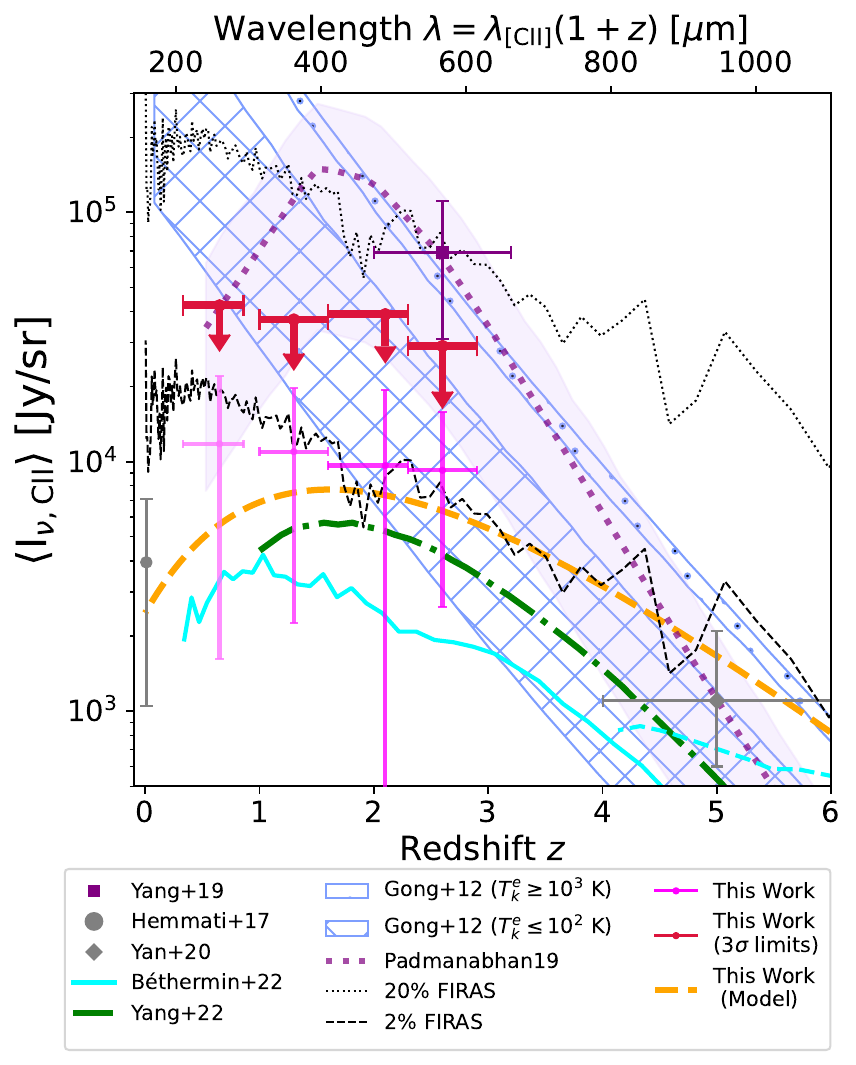}
\caption{
\reviewEditOne{Measurements of \ciium at $z\sim 0.3-2.9$ compared with \ciiumns-LF estimates in the local universe \reviewEditTwo{\citep{2017Hemmati}} and $z\sim5$ \reviewEditTwo{\citep{Yan_2020}}. Also shown are theoretical predictions from C+ evolution models, \reviewEditTwo{including \citet{2012Gong} (blue hatches), \citet{Padmanabhan_2019} (dotted purple: best model, shaded band: uncertainty), \citet{Yang_2022} (dash-dotted green), and \citet{bethermin+22} (cyan line: \citet{delooze+14} version; cyan dashed: high SFRD version at high $z$}. Our $3\sigma$ upper limits disfavor high-temperature collisional excitation frameworks and best-fit empirical models calibrated to the \citet{Pullen_2018, Yang_2019} \textit{Planck} measurement. The $1\sigma$ results are more consistent with SFR-scaling models, which calibrate C+ luminosity to the SFR of sources. 
Additionally, we show the \textit{COBE}/FIRAS measurement of the monopole spectrum of the CIB as a function of wavelength, matched to the rest-frame redshift of \ciium emission. The \ciium contribution is likely no more than a few percent of the total CIB, with the intensity history evolving by less than an order of magnitude across cosmic moon.
}}
\label{fig:constraints}
\end{figure}

\begin{figure}
\centering
\includegraphics[width=\linewidth]{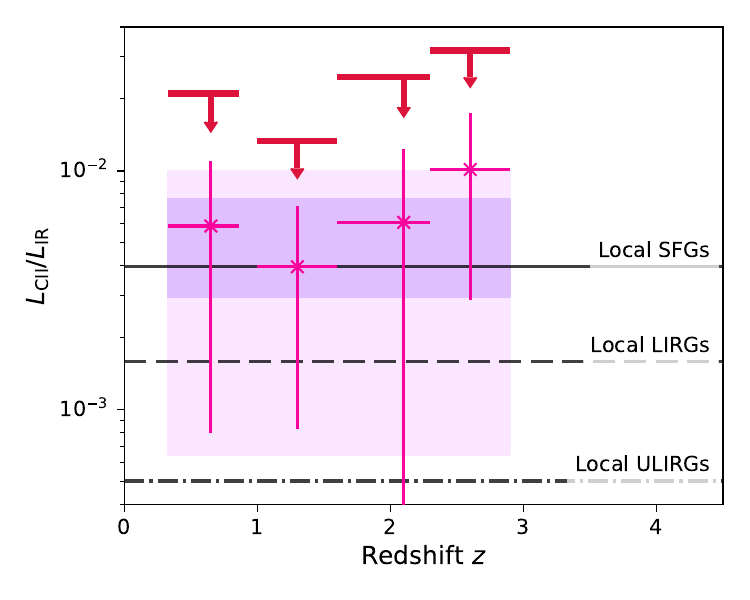}
\caption{
\reviewEditOne{Population-average \ciins-to-IR luminosity ratios, derived comparing our \cii constraints with integrated IR luminosity function of \citep{gruppioni+13}. Red limits and pink crosses indicate $3\sigma$ upper bounds and $1\sigma$ intervals in each redshift bin. Purple bars show the variance-weighted average over all bins ($z\sim0.33-2.9$), with dark purple and light purple indicating $1\sigma$ and $2\sigma$ intervals. We also show the average ratios for local SFGs from \cii observations from \citet{accurso+17}, and for LIRGs and ULIRGs from \citet{diaz-santos+17}.}
}
\label{fig:lciilir}
\end{figure}

We present our \ciium measurements in four tomographic bins in Figure \ref{fig:constraints} and Table \ref{tab:constraints}. Our constraints are inconsistent with no emission in three of these bins at $1.2-1.4\;\sigma$. We do not obtain a statistically non-zero detection at $z\sim2.1$; our 3$\sigma$ upper bound is plotted in red in Figure \ref{fig:constraints}. Overall, the quadrature sum of S/N in our four measurements is $\sim 2.42$. These constraints indicate that \ciium constitutes \reviewEditOne{no more than $4-8\%$} of the CIB (Figure \ref{fig:constraints} shows the \textit{COBE}/FIRAS measurement of the monopole CIB spectrum) at the corresponding observed wavelengths ($240-600~\mu$m).

We also plot \reviewEditOne{estimates} of \ciium at $z\sim 0$ and $z\sim6$, as well as theoretical predictions from selected C+ models in Figure \ref{fig:constraints}. 
\reviewEditOne{\citet{Yan_2020} combined targeted and serendipitous \cii detections with IR luminosity functions from the ALPINE survey to estimate the \cii luminosity function and mean intensity at $z\sim5$. \citet{2017Hemmati} estimated the local \cii luminosity function by applying scaling \reviewEditTwo{relations, derived from the GOALS\footnote{Great Observatories All-sky LIRG Survey} survey of local (ultra) luminous IR galaxies ((U)LIRGs), to a flux-limited sample} of IR-selected galaxies from IRAS\footnote{Infrared Astronomical Satellite}, using direct \cii measurements in place of estimates for the brightest galaxies. We integrated their double-power-law fit to the resulting luminosity function to derive an estimate of the local \cii luminosity density. We note that neither the $z\sim0$ nor $z\sim5$ measurements are based on a direct measurement of the \cii luminosity function (LF) and may have significant systematic uncertainties. Nevertheless, we include them as useful indications of the behavior of the mean \cii intensity outside the redshift interval we measured directly.}

\reviewEditOne{
To compare our results with theoretical expectations, we replicated predictions for the redshift evolution of \ciium from the literature. \citet{2021Yang, Yang_2022} (Y22) calibrate the galaxy ISM line emission luminosity versus halo mass relations to semi-analytical simulations. We reproduce the best model and uncertainties from the \citet{Padmanabhan_2019} (Pa19) prescription, which is designed to simultaneously reproduce the \citet{2017Hemmati} and \citet{Pullen_2018} measurements. Finally, we included a suite of models from \citet{2012Gong} (G12), which assume that C+ is proportional to the total carbon mass in dark matter halos. The G12 models treat electron kinetic temperature $T_k^e$ and number density $n_e$ as free parameters and list several models for different pairs; we plotted their range of estimates for their high temperature ($\log_{10}(T_k^e / \text{1 K}) = 3-4, \log_{10}(n_e/ \text{1 cm$^{-3}$}) = 2-4$) and low temperature ($\log_{10}(T_k^e / \text{1 K}) = 2, \log_{10}(n_e/ \text{1 cm$^{-3}$}) = 0-2$) models separately.
We also replicated the predicted \ciium intensity history from the Simulated Infrared Dusty Extragalactic Sky (SIDES) model \citep{bethermin+22} (B22), specifically showing the \citet{delooze+14} (D14) version with the ``high SFRD" variant (shown dashed) at high redshifts.
}

\reviewEditOne{
Finally, we implemented our own SFR-tracing model using the \textsc{SimIM} modeling framework \footnote{see \href{https://simim.readthedocs.io/latest/}{simim.readthedocs.io}}. 
We briefly describe the methodology and refer to \citet{2020ApJ...901..141K, Keenan_2022, Keenan_2020, keenan} for additional details. Starting from the magneto-dynamic simulation suite IllustriusTNG, specifically \texttt{TNG100-1}, \citep{2018MNRAS.477.1206N, 2018MNRAS.475..676S, 2018MNRAS.475..648P, 2018MNRAS.475..624N, 2018MNRAS.480.5113M}, for each volume at a given redshift, we prescribed a subgrid line emission model, from empirical calibration of subhalo mass to SFR \citep{2013Behroozi} and constraints on the scaling between \cii luminosity and SFR \citep{delooze+14}. 
Finally, we applied a constant amplitude scaling to best fit the constraints from \citet{2017Hemmati}, \citet{Yan_2020}, and this analysis. 
}

\reviewEditOne{
The empirical B22 SIDES-based model constructed above and the Y22 semi-analytical models imply a close connection between \cii and star formation, and we designate them broadly as ``SF-tracing.'' Conversely, the G12 model requires no connection between \cii and SFR, while the Pa19 model relies on a more-than-linear redshift evolution of \ciins-SFR correlation to reproduce the \citet{Pullen_2018} data point. 
Our $3\sigma$ limits are inconsistent with the best-fit Pa19 model and with all high-temperature versions of the G12 prescription at all $z$. We excluded the low-temperature G12 models at $3\sigma$ in our lowest redshift bin, $z\sim0.65$, with some models falling within our $1\sigma$ bar at higher redshifts. Lower-intensity versions within the uncertainty range of the Pa19 model space are also within the $3\sigma$ limits. The predictions of the SF-tracing models lie well within the upper limits and are consistent within the $1\sigma$ uncertainty bars. 
Our measurements therefore disfavor scenarios in which the ratio between \cii luminosity and SFR evolves dramatically 
\reviewEditOne{across cosmic noon}. Our results from this untargeted survey are consistent with targeted studies that have found near-linear scaling between \cii luminosity and SFR with little to no redshift evolution \citep{delooze+14,schaerer+20,2022Romano}, although higher sensitivities would be required for a robust confirmation.
}

\begin{table*}[t]
\caption{Estimated bias from contaminating spectral lines within the \textit{Herschel}/SPIRE broad bandpass for each \ciium redshift bin. \label{tab:contaminants}}
\begin{tabular}{c|c|c|c|c}
\centering
$z_\text{median}$ & $(z_{\text{min}}, z_{\text{max}})$ & SPIRE band & Contaminating Interlopers within filter FWHM & 
$\frac{\sum_\text{inter}(L \tilde{T})}{L_\text{\cii} \tilde{T}_\text{\cii} + \sum_\text{inter}(L \tilde{T})}$ 
\\\hline
.65 & (.33, .86) & 250$\mu$m & \makecell{[OI]-145$\mu$m, [NII]-122$\mu$m, $^{12}$CO(12-11),  \\ $^{12}$CO(13-12), [NII]-205$\mu$m} & 
6.8\%
\\
1.3 & (1.0, 1.6) & 350$\mu$m & [OI]-145$\mu$m, [NII]-122$\mu$m, $^{12}$CO(13-12) & 
6.5\%
\\
2.1 & (1.6, 2.3) & 500$\mu$m & [OI]-145$\mu$m, $^{12}$CO(12-11), $^{12}$CO(13-12), [NII]-205$\mu$m & 
7.3\%
\\
2.6 & (2.3, 2.9) & 500$\mu$m & [OI]-145$\mu$m, [NII]-122$\mu$m & 
11\%
\end{tabular}
\\
{Notes:} {Estimates of line luminosity are taken from \citet{Visbal_2011}; we consider all lines in their Table 1. We identify interlopers redshifted within the FWHM of the SPIRE bands and report the aggregate potential positive bias. These biases are insignificant compared with our uncertainties.} 
\end{table*}

\reviewEditOne{
In the nearby Universe, (U)LIRGS exhibit low $L_\text{\cii} / L_\text{IR}$ ratios \reviewEditTwo{($L_\text{\cii}$ and $L_\text{IR}$ being the \ciium line luminosity and the IR luminosity, respectively)}, relative to non-IR-selected samples \citep{2017ApJ...846...32D}. At cosmic noon, around half of star formation occurs in (U)LIRGS \reviewEditTwo{\citep{Casey_2014}}, implying that a significant fraction of the mean \ciium intensity must also originate in such systems. To check for evidence that galaxies contributing significantly to $I_\nu^\text{\cii}$ at $0.33<z<2.9$ are [CII] deficient, we computed} 
the ratio of the integrated \cii and total IR (TIR) luminosities in each redshift bin. To achieve this, we calculated the luminosity-weighted integral of the IR luminosity functions of \citet{gruppioni+13} \footnote{\reviewEditOne{Other luminosity functions could affect this discussion; we compared our lowest $z$ bin with the TIR LF from 
\citet{2013MNRAS.428..291P} and our lowest three $z$ bins with the TIR LFs from \citet{2011A&A...528A..35M}; in all cases, we find consistent results.}} at the central redshift of each bin and computed the ratio with $I^{\text{\ciins}}_\nu(z)$ (converted into units of L$_\odot$ Mpc$^{-3}$). Figure~\ref{fig:lciilir} compares our constraints on $L_{\text{\ciins}}/L_{\rm IR}$ with the ratios for \reviewEditOne{local SFGs and (U)LIRGS observed by Herschel, derived from \citep{accurso+17} and \citep{diaz-santos+17}, respectively}. Taking the variance-weighted average of all \reviewEditOne{redshift} bins yields $L_{\text{\ciins}}/L_{\rm IR}=0.0053\pm0.0023$, which is very similar to the value observed in SFGs in the nearby Universe. Our constraint lies 2$\sigma$ above the median ratio seen in (U)LIRGs. \reviewEditOne{This suggests} that the galaxies that make up the bulk of the \cii emission at cosmic noon do not exhibit large \cii deficits; however, confirmation with more sensitive data is required. \reviewEditOne{
While our data only constrain the $L_\text{\cii} / L_\text{IR}$ ratios for galaxies responsible for the bulk of FIR emission at these redshifts, our findings are consistent with \citet{2018MNRAS.481.1976Z}, who show that $z\sim2$ SFGs do not exhibit deficient $L_\text{\cii} / L_\text{IR}$ ratios, despite their (U)LIRG-like total IR luminosities.}

\subsection{\reviewEditOne{Contaminating spectral lines} \label{subsec:contamination}}

\reviewEditOne{
\ciium is not the only narrow non-continuum spectral feature in these bands, and other emission lines can originate from (and hence correlate with) galaxies within our specified $z$ bins. Thus, some of the excess emission we measure over the continuum could originate from lines other than \ciiumns.}

\reviewEditOne{
To estimate this positive bias on our \ciium measurement, we considered all emission lines listed in Table 1 of \citet{Visbal_2011}. |The table lists the rest-frame wavelengths of each line, along with a calibration ratio between line luminosity and SFR $L_\text{line}/\dot{M_*} = R$. We assume that \reviewEditTwo{the ratio of the luminosities of the \ciium and the interloper line (denoted by ``inter'')} is given by the ratio of their luminosity-SFR calibration factors,
\begin{align}
L_\text{inter} \Big/ L_\text{\cii} = R_\text{inter} \Big/ R_\text{\cii}.
\end{align}
\citet{Pullen_2018} also estimate interloper bias on their measurements using the same table under a similar assumption for line luminosity scaling. 
}

Next, given the transmission $T(\lambda)$ of each SPIRE map's broadband filter (see Fig. \ref{fig:filter_curves}), we computed the effective transmission coefficient for each contaminant line within each redshift bin in Table \ref{tab:constraints}. Contaminant emission from interloper lines in our measurement is thus \footnote{There is an implicit assumption of uniform emitter density in $z$ within each bin in Eqn. \ref{eqn:inter}; for brevity, we ignored this effect and did not expect it to significantly change the values in Table \ref{tab:contaminants}.}
\begin{align}
\label{eqn:inter}
I_\text{inter}^\text{contaminant} \propto L_\text{inter} \times \tilde{T}_\text{inter} \propto R_\text{inter} \int_{\lambda(z_\text{min})}^{\lambda(z_\text{max}) }T(\lambda) d\lambda,
\end{align}
\reviewEditOne{
where the two terms respectively encapsulate the proportional brightness of the line being constrained (\ciiumns) and the relative overlap of interloper line emission at the correct redshifts and wavelengths, which we optimized for in our binning (as discussed in Section \ref{subsec:simstack} and Fig. \ref{fig:filter_curves}). 
}

\reviewEditOne{
Table \ref{tab:contaminants} lists each of the bins in which we report a \cii constraint, the redshift ranges, lines from the collection in \citet{Visbal_2011} that are redshifted into the FWHM of the corresponding SPIRE map, and finally, an estimate of the fraction of the excess emission we measure that could be attributed to emission from non-\ciium lines. The main contaminants are [OI]-145$\mu$m, as it is very close to the rest-frame wavelength of the \cii line, and [NII]-122$\mu$m, whose overall line luminosity is expected to be within an order of magnitude of \ciiumns \citet{Visbal_2011}). Overall, we expect contamination from interlopers to be within $6.5-11\%$ of the measured total excess emission; we consider the effects on our reported ($\sim1-1.4\sigma$ significance) results subdominant.
}

\reviewEditOne{
Mapping FIR fields at higher spectral resolution would be the obvious way forward to address contaminants, lowering the effective transmission, $\tilde{T}$, for interlopers in Eqn. \ref{eqn:inter}.
It is noteworthy for the discussion in Section \ref{subsec:tension} that correcting for the effect of contaminants would lower our result, increasing the tension with the \textit{Planck}$\times$BOSS results.
}

\subsection{\reviewEditOne{Other systematics} \label{subsec:systematics}}

\reviewEditOne{
We briefly discuss other potential systematics in our best estimates and in the $3\sigma$ upper limits.
}

\paragraph{Photometric redshift uncertainty:}
The COSMOS2020 photometric galaxy catalog claims sub-percent accuracy for \reviewEditTwo{bright objects in the $i$-band} ($i<21$) and $<5$\% precision for the faintest ($25<i<27$) sources \reviewEditTwo{\citep{weaver_cosmos2020_2022}}. As our broadband filters and corresponding $z$-bins are wide, we do not expect photo-$z$ errors to significantly and systematically move galaxies with \ciium emission into neighboring bins.
To quantify the effects of $z$ errors, we generated the same catalog used in the analysis above, except sampling the estimated photometric $z$ from a skewed Gaussian distribution that replicates the median, lower 68\%, and upper 68\% of the probability distribution functions ($z$-PDFs) in the COSMOS2020 catalog. We reran our stacking and SED fitting analysis on this catalog with the modified sampled redshifts. We find that our \ciium constraints deviate by no more than 0.09$\sigma$ in any of the four $z$ bins, which we consider sub-dominant to our measurements.

\paragraph{Diffuse emission from intergalactic medium and unmodeled extragalactic background light:} Our stacking methodology measures \ciium emission from physical scales within the \textit{Herschel} beam of the sources in the COSMOS2020 catalog, which is $\sim 0.2-1$ $c$Mpc at $z\sim 0.3-2.9$. We expect this to account for the circumgalactic medium and the diffuse \cii halo around sources, which are typically within $\lesssim100$ kpc. \citep{2017ARA&A..55..389T, 2020ApJ...900....1F,2023MNRAS.518.3183L,2019ApJ...887..107F,2025A&A...693A.237I,2015ApJ...800....1H}
Our results could be negatively biased due to underestimating emission from the large-scale diffuse intergalactic medium (IGM). However, this effect is subdominant, as the mean \ciium emission from the IGM is expected to be orders of magnitude lower than galaxies \citep{2012Gong}.

\reviewEditOne{\reviewEditTwo{The emission in the \textit{Herschel}/SPIRE bands can be resolved entirely (within error bars) via source stacking in the COSMOS field \citep{2020Duivenvoorden}}. However, notably, their catalog is non-tomographic, and we can only stack the subset with photo-$z$ estimates in this analysis. 
Section \ref{sec:completeness} describes the corrections we applied to account for the unmodeled background in the FIR, which included modifications based on stellar mass function (SMF) measurements, as well as a constant scaling to match the COBE/FIRAS monopole spectrum. Invalidity of the assumptions underlying these two extrapolations would bias our measurement. 
Potentially more complete photo-$z$ catalogs exist, such as COSMOS2025 \citep{shuntov2025cosmos2025cosmoswebgalaxycatalog}, but they cover smaller portions of the sky, resulting in reduced sensitivity when stacking. 
}

\subsection{\reviewEditOne{Tension with past constraints at $z\sim 2.6$} \label{subsec:tension}}

\begin{figure}[ht]
\centering
\includegraphics[width=\linewidth]{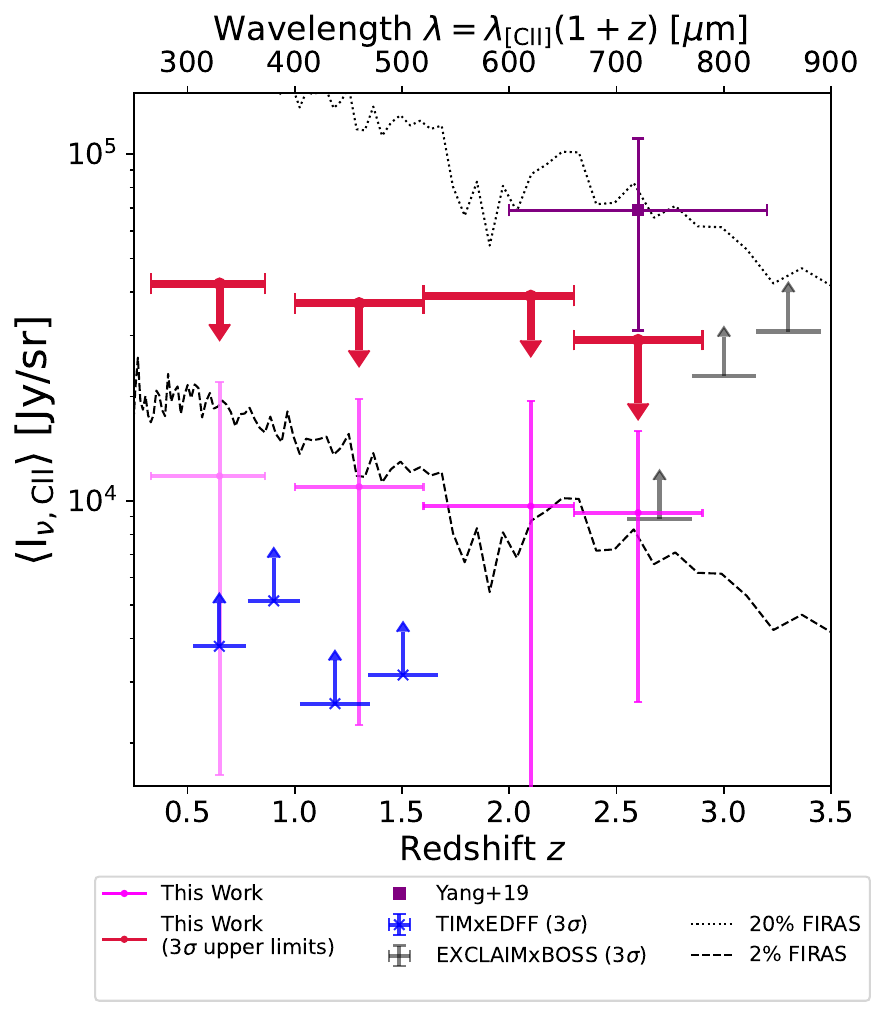}
\caption{Constraints with $3\sigma$ upper bounds vs. 95\% C.L. bounds from \citet{Yang_2019} (a followup of \citet{Pullen_2018}) measurement at $z\sim 2.6$. No \ciium is detected at the previously tentatively reported levels; our best estimates are $\sim7.5\times$ lower, with the other $3\sigma$ upper bound at $z\sim2.6$ lower than the 95\% C.L. interval.
Expected $3\sigma$ sensitivity limits of two in-development surveys \citep{Pullen_2023, timconstraints}), which will measure C+ at high $R$, are also shown. This tension is discussed further in the text (Section \ref{sec:results}).}
\label{fig:compare}
\end{figure}

In Figure \ref{fig:compare}, we directly compare our measurements with the \citet{Pullen_2018, Yang_2019} constraint at $z\sim 2.6$. In their analyses, Y19+P18 cross-correlate quasars and CMASS galaxies from SDSS BOSS with three high-frequency IR all-sky maps from the \textit{Planck} mission. This enables a search for excess emission in the middle 545 GHz band, with the cross-power spectrum in the 353–857 GHz bands constraining nuisance parameters, including the CIB and the thermal Sunyaev-Zeldovich contribution. This excess emission is theorized to be \ciiumns, expected to be the brightest line feature in the band.
Previous measurements by Y19+P18 indicate $I^{\text{\ciins}}_\nu \sim 69^{+42}_{-38}$ $k$Jy/sr at 95\% C.L., amounting to $\sim10-30\%$ of the CIB at 545 GHz. Our measurement in the $z\sim2.6$ tomographic bin is $\sim7.5\times$ smaller and inconsistent at about \reviewEditOne{$2.9\sigma$}. Notably, our 3$\sigma$ upper bound at 29 $k$Jy/sr is still below the 95\% confidence lower limit reported by Y19+P18  

Assuming the validity of either analysis, we discuss the physics or systematics that could explain this \reviewEditOne{$2.9\sigma$} tension and inconsistency of nearly an order of magnitude in the \ciium intensity history just prior to the peak of cosmic star formation. 
The Y19+P18 studies cross-correlate all-sky fields, utilizing scales as large as \reviewEditOne{multipole moment $l\sim 100$ (or angular scales of $\sim 1\deg$)}. Therefore, they explore cross-correlation at larger scales than the \textit{Herschel}/SPIRE beam at $\sim 500~\mu$m, which corresponds to $\sim1$ $c$Mpc at $z\sim2.6$. \reviewEditOne{As discussed in Section \ref{subsec:systematics}}, if C+ predominantly exists outside of $\sim1$ $c$Mpc of sources, our analysis would underestimate \ciiumns, but line emission would still correlate at large scales in \textit{Planck} maps and Y19+P18 would detect it.
\reviewEditOne{
However, \citet{2012Gong} models the IGM \ciium contribution to be orders of magnitude lower than that of galaxies.
}

Our analysis uses a photometric catalog that is incomplete; as discussed in Section \ref{sec:completeness}, we measured and attempted to correct for this incompleteness. We scaled our \ciium measurements by the inverse of the fraction of the COBE/FIRAS CIB monopole spectrum that we resolve; this assumes an underlying extrapolation that the same fraction of the emission from sources not in our photometric catalog is \ciins. Our measurements would therefore be an underestimation if sources missing from the COSMOS2020 photometric redshift catalog were anomalously bright emitters of \ciiumns. However, it is worth pointing out that we only miss $\sim 15-25\%$ of the CIB in the SPIRE bands; thus, a significant portion of the IR emission from the missing sources would have to be \cii to account for the tension with Y19+P18, which posits $\sim 20\%$ of the CIB to be \ciiumns. 

Examining the final column of Figure \ref{fig:sed_fits} (or Figure 7 from \citet{Pullen_2018}), we note that Y19+P18 uses three points, all $>350~\mu$m, to constrain a similar continuum SED model as our analysis; empirically, using a low number of samples only on the Rayleigh-Jean side of the modified black-body model could be prone to fitting and numerical systematics. More data points at 
wavelengths away from the bands in which \cii\ emission would
enable an analysis like ours to robustly constrain the nuisance SED parameters.

\subsection{\reviewEditOne{Future outlook} \label{subsec:future_outlook}}

Figure \ref{fig:compare} also presents sensitivity forecasts for two in-development instruments. We include the $3\sigma$ noise limits expected from cross-correlating datasets from two C+ spectrometers with a respective ancillary galaxy catalog. The Terahertz Intensity Mapper (TIM; \citet{vieira2020tim}) is a balloon-borne FIR spectrometer, \reviewEditOne{with spectral resolution $R\sim250$}, mapping a deep field at GOODS-South for $O(100)$ hours. The field observed by TIM overlaps with the Euclid Deep Field Fornax (EDFF); TIM plans to cross-correlate their \ciium maps at $z\sim0.5-1.5$ with \reviewEditOne{the} deep high number-count \textit{Euclid} spectroscopic galaxy catalog \citep{timconstraints}. \reviewEditOne{The EXperiment for Cryogenic Large-Aperture Intensity Mapping (EXCLAIM) \citep{Pullen_2023} will map \ciium at $z\sim2.5-3.5$ over an equatorial field} for $\sim10$ hours. The experiment plans to cross-correlate with the SDSS BOSS spectroscopic redshift catalog. Figure \ref{fig:compare} shows that if either experiment is able to reach its predicted sensitivity, it will be a step towards resolving this perceived tension by detecting \cii at several $\sigma$. 

Constraints from our methodology can be improved by increasing either the depth of the FIR broadband maps, the number counts in the galaxy catalog, \reviewEditOne{or by addressing systematics with smaller photo-$z$ uncertainty or narrower bandpass measurements}. We are currently limited to broadband mapping instruments of the past, such as \textit{Herschel} and\textit{Spitzer}), but wider and/or more complete galaxy catalogs are imminent. Deeper UV-O-IR measurements of COSMOS, such as with \textit{JWST}/NIRCam or \textit{JWST}/MIRI \citep{2023ApJ...954...31C,shuntov2025cosmos2025cosmoswebgalaxycatalog}, could improve the FIR completeness of our forward model. More promisingly, galaxy catalogs with the same depth but more overlap with the \textit{Herschel} deep fields, e.g., \textit{Euclid} photometry and spectra \citep{EuclidOverview, 2025Euclid}, would decrease our constraint variance as roughly inversely proportional to the total number counts of emitters with photo-$z$ in the coinciding regions. 

\section{Conclusions \label{sec:conclusion}}

We place the first constraints on \ciium emission during and immediately after the peak of star formation history, at $z\sim0.6-2$. These results imply that  \ciium emission increased by roughly an order of magnitude from $z\sim 5$ and then fell by a similar amount to the present day, accounting for no more than a few percent of the cosmic IR background at its peak. Our measurements, with a total S/N of 2.42, imply $I^{\text{\ciins}}_\nu$ of $\sim10$ kJy/sr across $z\sim0.6-2.6$.

We are unable to replicate measurements at $z\sim 2.6$ obtained from cross-correlating \textit{Planck} IR maps with quasars from the wide spectroscopic survey BOSS. Experiments currently in development, TIM and EXCLAIM, are forecasted to be sensitive enough to address this tension and measure \cii at several $\sigma$s. Additionally, imminent data releases of the deep-field galaxy catalogs from the recently launched \textit{Euclid} mission, as well as the COSMOS-Web survey, are expected to yield higher-completeness coverage over larger fields overlapping with \textit{Herschel} deep maps. A future extension of our analysis with \textit{Euclid} redshifts could leverage such higher effective number counts to decrease uncertainties on stacked fluxes, and thus $I^{\text{\ciins}}_\nu$.

We have demonstrated a search for FIR line emission in deep field multi-wavelength surveys, using \linsimstackns, a faster linearized version of previously introduced stacking utilities. It is straightforward to extend our analysis to obtain complementary constraints from different datasets, such as \textit{Euclid}, and from different fields, such as GOODS. \reviewEditOne{Our methodology can also be used to constrain the redshift evolution of molecular lines from CO, \reviewEditOne{[OI], [OIII], and} N+, provided that these narrow spectral features are expected to be significantly brighter than contaminating interlopers, or that maps at higher spectral resolution are available.}

\begin{acknowledgements}
We thank \reviewEditOne{Garrett} Keating, \reviewEditOne{Adam Lidz,} {Ian Smail}, Joaquin Vieira, and Jessica Zebrowski for insightful discussion. \reviewEditOne{We thank the anonymous reviewer for their comments.} The S2CLS JCMT data used in this paper were taken as part of Program ID MJLSC02.

This publication was made possible through the support of Grant 63040 from the John Templeton Foundation. The opinions expressed in this publication are those of the authors and do not necessarily reflect the views of the John Templeton Foundation. Shubh Agrawal's work was also supported by the Quad Fellowship.
\end{acknowledgements}

\vspace{5mm}
\textit{Facilities: }{COSMOS2020 \citep{weaver_cosmos2020_2022}: GALEX, MegaCam/CFHT, ACS/\textit{HST}, HSC/Subaru, Suprime-Cam/Subaru, VIRCAM/VISTA, IRAC/\textit{Spitzer}; Maps: MIPS/\textit{Spitzer}, SPIRE/\textit{Herschel}, PACS/\textit{Herschel}, SCUBA2/JCMT.}

\textit{Software: }{astropy \citep{2013A&A...558A..33A,2018AJ....156..123A}; emcee \citep{Foreman-Mackey_2013}, \linsimstackns, an fork of SIMSTACK3 \citep{Viero_2022, 2013Viero}}

\bibliography{AA56503-25}{}
\bibliographystyle{aa}

\newpage

\appendix
\section{\linsimstack Stacking outputs for the three \textit{Herschel}/SPIRE maps \label{app:linsimstack}}

\begin{figure}[hbtp]
\centering
\includegraphics[width=\linewidth]{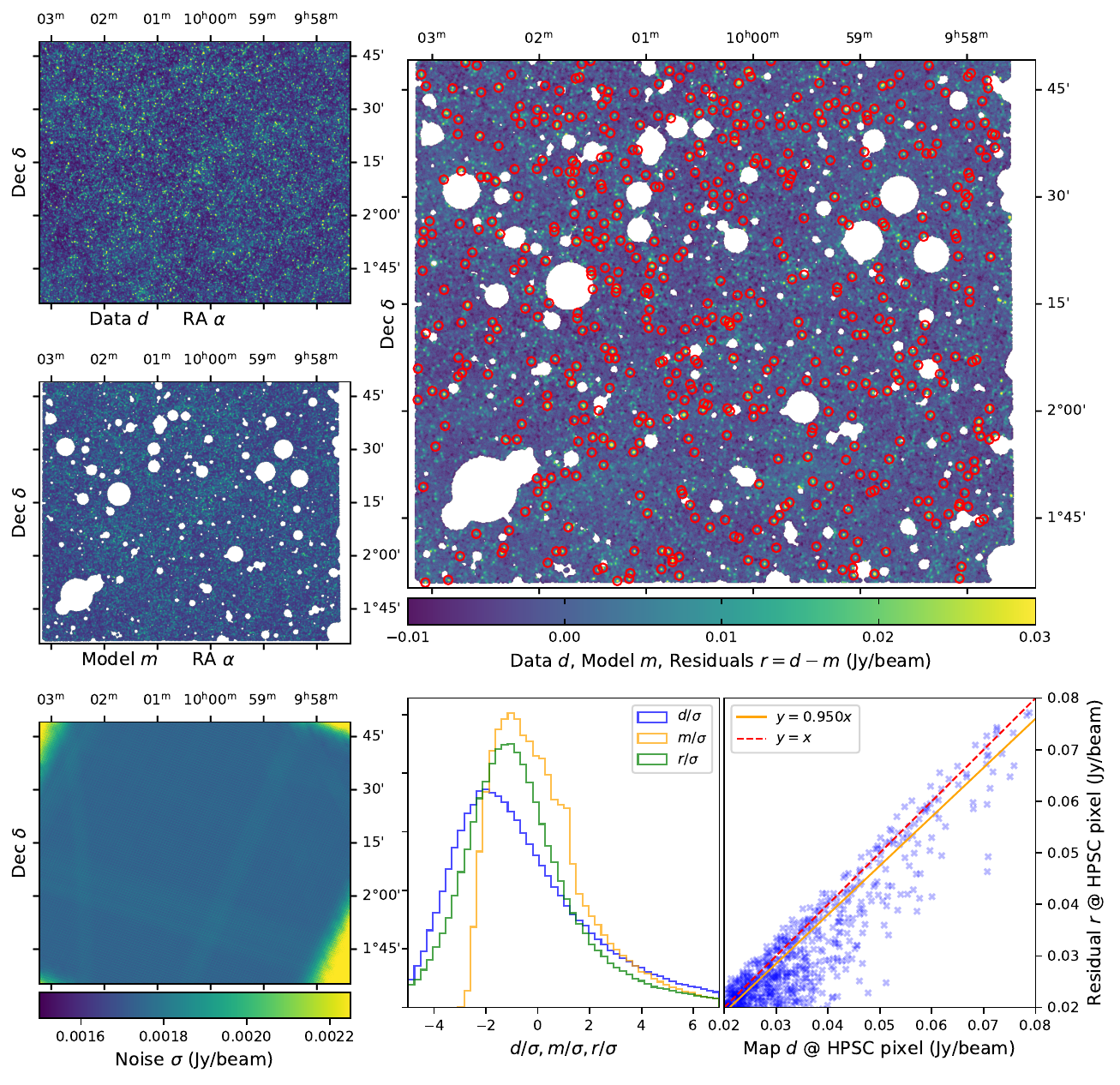}
\caption{\textit{Left Top:} \textit{Herschel}/SPIRE 250 $\mu$m map of the COSMOS field, used in the forward modeling framework of \linsimstack as the data vector $d$. This corresponds to our $z\sim0.65$ tomographic bin for \ciiumns.\\
\textit{Left Middle:} Model $m$ from \linsimstackns, a linear combination of hit-maps of individual bins, convolved with the 250 $\mu$m beam, with the coverage mask of COSMOS2020 applied. \\
\textit{Top Right:} Residuals $d-m$ not resolved by \linsimstack. We overplot the positions of the brightest 500 COSMOS sources in the Herschel SPIRE Point Source Catalog at 250 $\mu$m. \\
\textit{Bottom Left:} Per-pixel noise in the 250 $\mu$m map. \\
\textit{Bottom middle:} Histograms of the data, \linsimstack fit, and the unresolved residuals, all scaled by the per-pixel noise level after masking. Each is mean-subtracted independently. The residuals are more (standard) Gaussian than the map, with a high-end tail noting the missing sources in the COSMOS2020 catalog. \\
\textit{Bottom right:} Flux in the map versus in the residuals at the pixel location of sources in the \textit{Herschel}/SPIRE point source catalog. Also plotted is a linear fit, with each pixel location weighted by the SNR; this encapsulates a measure of how much \linsimstack under-resolves the HSPSC sources.}
\label{fig:SPIRE_PSW}
\end{figure}

\begin{figure}[hbtp]
\centering
\includegraphics[width=\linewidth]{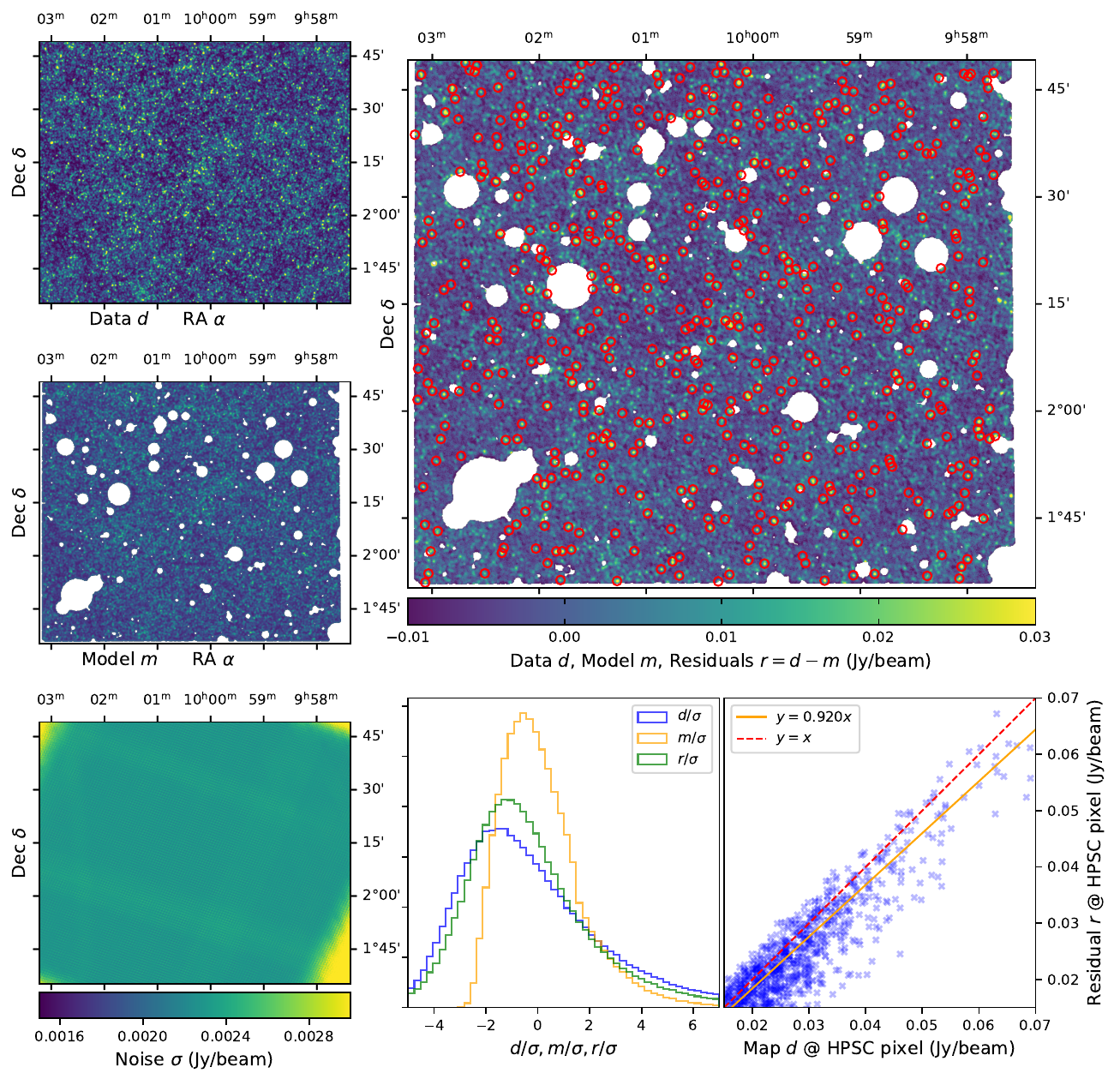}
\caption{Same as Figure \ref{fig:SPIRE_PSW}, but for the \textit{Herschel}/SPIRE 350 $\mu$m map. This corresponds to our $z\sim1.3$ tomographic bin for \ciiumns.}
\label{fig:SPIRE_PMW}
\end{figure}

\begin{figure}[htp]
\centering
\includegraphics[width=\linewidth]{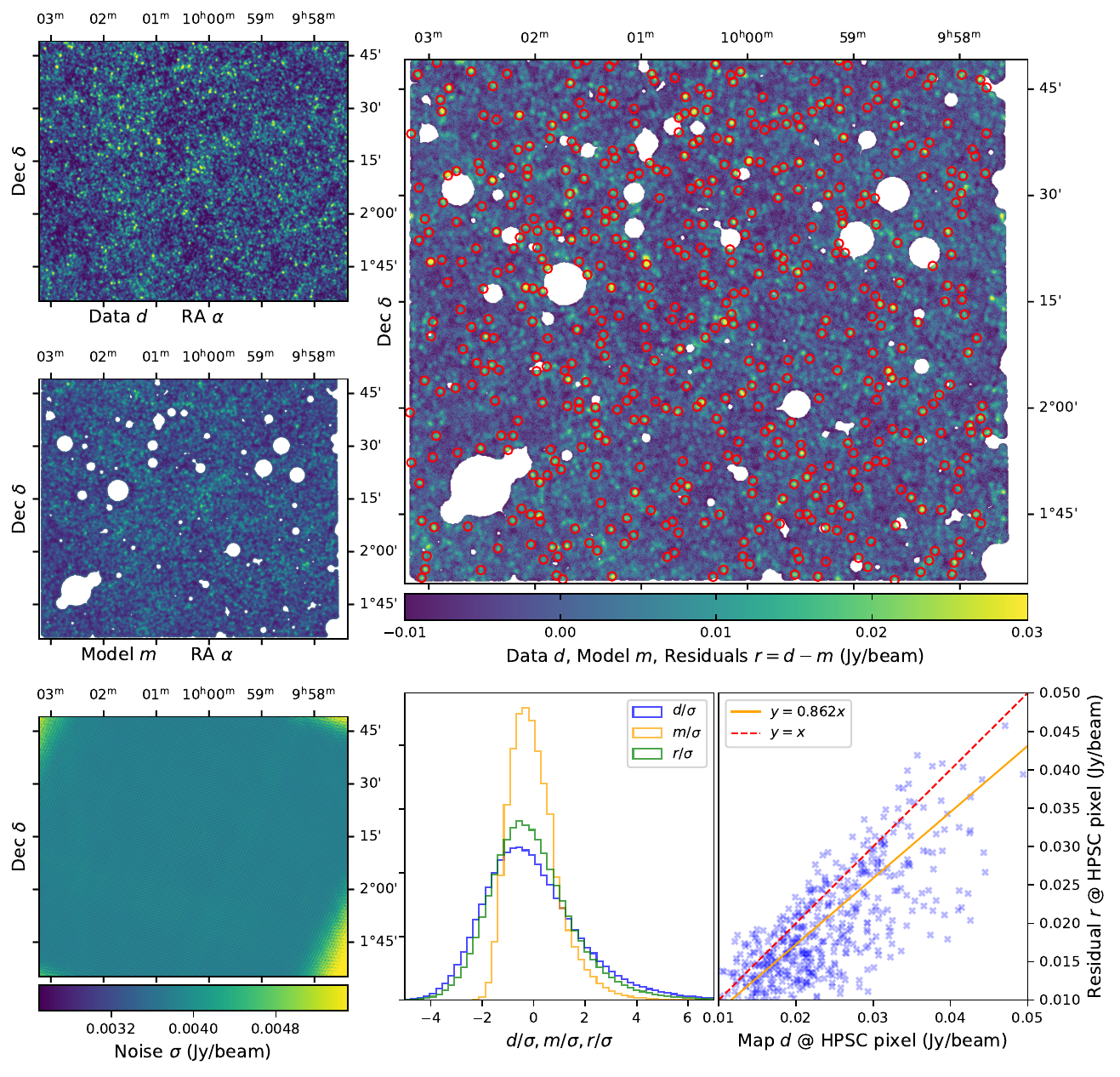}
\caption{Same as Figure \ref{fig:SPIRE_PSW}, but for the \textit{Herschel}/SPIRE 500 $\mu$m map. This corresponds to our $z\sim2.1$ and $z\sim2.6$ tomographic bins for \ciiumns.}
\label{fig:SPIRE_PLW}
\end{figure}

Figures \ref{fig:SPIRE_PSW}, \ref{fig:SPIRE_PMW}, and \ref{fig:SPIRE_PLW} present an overview of the stacking-based forward modeling of the \textit{Herschel}/SPIRE maps. In Section \ref{sec:completeness}, we used the \textit{COBE}/FIRAS monopole measurement to quantify the incompleteness of our photometric redshift catalog and found that we fail to resolve $\sim13\%$, $\sim24\%$, and $\sim27\%$ of the CIB in the SPIRE bands. We empirically noted that the brightest sources in the 2D data vector $d$ were missing in the model vector $m$. Sources from the \textit{Herschel}/SPIRE Point Source Catalog (HSPSC) \footnote{\href{https://doi.org/10.5270/esa-6gfkpzh}{doi.org/10.5270/esa-6gfkpzh}} correlate with the residual map, seen in \textit{top right} and \textit{bottom right} figures; \linsimstack does not resolve $86\% - 95\%$ of the flux from the HSPSC. Based on flux estimates from the HSPSC, these sources sum up to 16\%, 12\%, and 10\% of the FIRAS CIB monopole spectrum in each of the three SPIRE bands. Either the COSMOS2020 catalog fails to detect or estimate a photometric $z$ for the object with UV-O-IR observations, or the HSPSC sources are anomalously brighter in the SPIRE bands than other sources with similar stellar mass contents at similar $z$.


\section{Stacking results \label{app:allstacking}}
\begin{figure*}[bt]
\centering
\includegraphics[width=\linewidth]{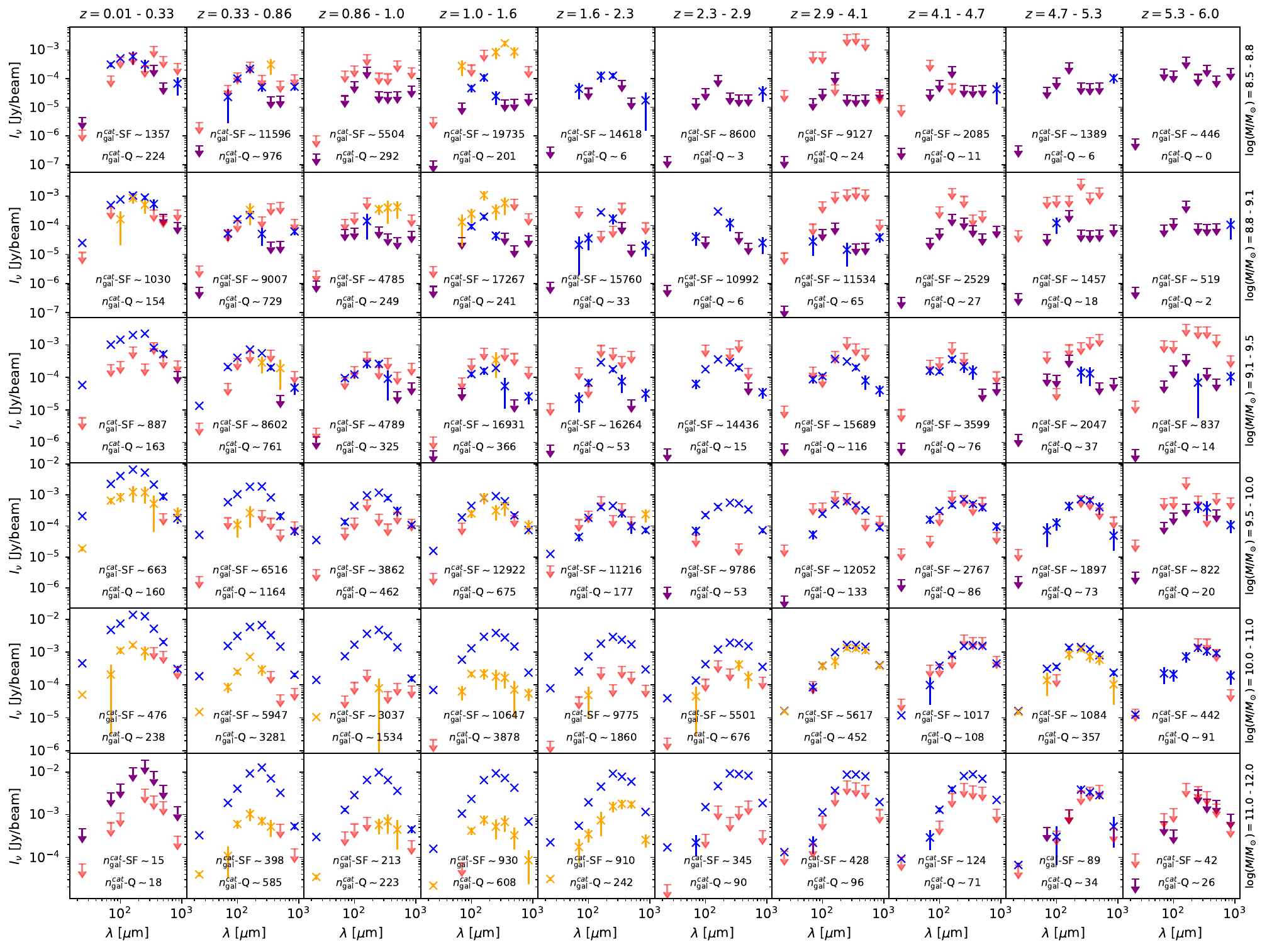}
\caption{All results obtained from \linsimstack, while stacking the COMSOS2020 photometric redshifts on 8 maps from the mid-IR to submillimeter. Detections in the star-forming and quiescent selections are denoted by blue and orange crosses, respectively, along with their $1\sigma$ uncertainties. For non-detections (measurements consistent with zero), the $3\sigma$ upper limits are shown as purple and red arrows for the star-forming and quiescent selection, respectively. Also annotated are the number counts for each selection with the redshift+stellar-mass bins in the photo-$z$ catalog. We do not include stacking estimates for number counts lower than 10; we empirically noted that the SED and bootstrapping errors could not be well constrained within these bins due to low number counts. }
\label{fig:allstacking}
\end{figure*}

\reviewEditOne{
Fig. \ref{fig:allstacking} shows the stacking outputs from \linsimstack for all of the COSMOS2020 bins, divided by star-forming (SF) versus quiescent (Q) (based on a color selection), redshift ($0.01<z<6$), and stellar mass content ($8.5<\log(M/M_\odot)<12$). All bins are \textit{simultaneously} stacked (see Section \ref{subsec:simstack}), including both the SF (blue and purple resp. for detections and non-detections) and the Q selections (yellow and red resp. for detections and non-detections). We marginalize residuals over all stellar mass bins and SF-Q selection, within a given redshift bin to get our \ciium constraints in Fig. \ref{fig:constraints}.}

\section{Comparing \linsimstack to SIMSTACK3 from \citet{Viero_2022}}

\begin{figure*}[bt]
\centering
\includegraphics[width=\linewidth]{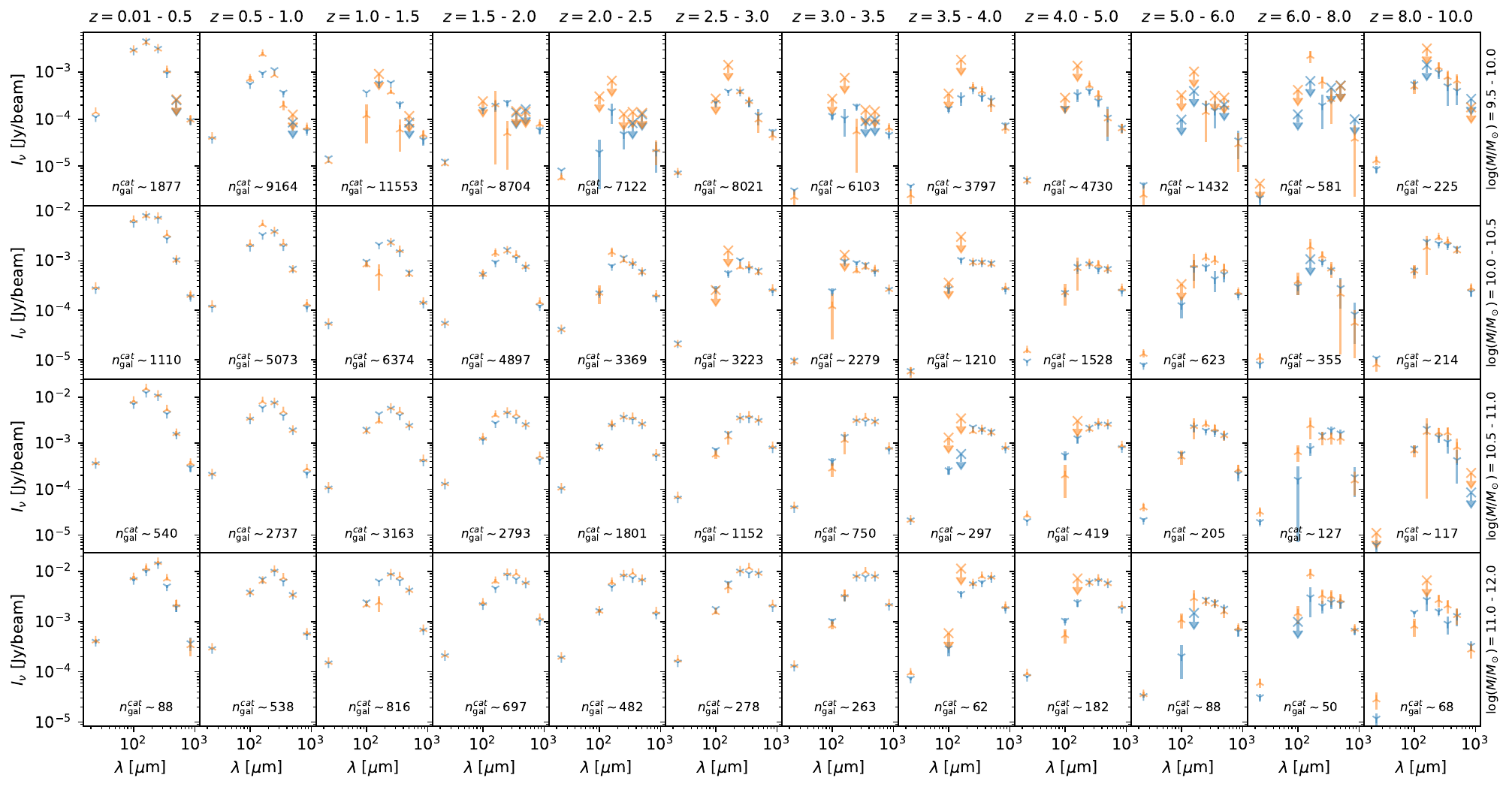}
\caption{Replicating Figure 1 from \citet{Viero_2022}, to compare results from \linsimstack (blue) to SIMSTACK3 (orange). Error bars plotted are from catalog bootstrapping alone (see Section \ref{subsec:errors}), to allow a direct comparison with SIMSTACK3/V+22. The RMS of $z$-scores is $\sim 1.35$, close to the expected unity.  We expect our methodology to be more robust against local minima in the loss function.}
\label{fig:viero_rep}
\end{figure*}

\reviewEditOne{
Our stacking methodology is similar to SIMSTACK3 from \citet{Viero_2022} (\reviewEditTwo{V+22}), with a crucial change. While SIMSTACK3 formulates the stacking as a non-linear problem and uses a Levenberg–Marquardt algorithm to iteratively converge to a best-fit solution, \linsimstack exploits the linearity of the forward modeling problem to gain an order-of-magnitude speed-up. Notably, our method, unlike SIMSTACK3, does not require a feasible initial guess to converge to the global minima. In Figure \ref{fig:viero_rep}, we attempt to replicate stacking results from V+22 (see their Figure 1), by using V+22's specific choice of binning over COSMOS2020. 
}

\reviewEditOne{
Results from \linsimstack (blue) typically agree with SIMSTACK3 results (orange); the RMS of $z$-scores is $\sim 1.35$, close to 1. Note that the uncertainties plotted and used in the calculation above are just catalog bootstrapping errors (which is all that V+22 use); if we used our more conservative error estimation (Section \ref{subsec:errors}), our uncertainties will be $\sim\sqrt{2}$ larger (see Fig. \ref{fig:errors}), resulting in a lower statistic above.
We expect \linsimstack is more robust against local extrema in the loss function; we qualitatively note that our stacking estimates trace the continuum better than SIMSTACK3. }

\section{Catalog and model predictions for source number counts in COSMOS2020}
\reviewEditOne{
Tables \ref{table:N_star} and \ref{table:N_nonstar} list number counts in the COSMOS2020 photo-$z$ catalog \citet{weaver_cosmos2020_2022} and those obtained from the stellar mass function (SMF) \citet{Weaver_2023} model, for the star-forming and quiescent selection of sources, respectively.
}

\begin{table*}[bt]
\scriptsize
\centering
\caption{\label{table:N_star} Number counts for the star-forming selection of galaxies in the COSMOS2020 photometric redshift catalog, used for simultaneous stacking with \linsimstackns.}
\centering
\begin{tabular}{c|cccccc}
\hline
\hspace{1cm}$\log_{10}(M/M_\odot) \in$ & (8.5, 8.8) & (8.8, 9.1) & (9.1, 9.5) & (9.5, 10.0) & (10.0, 11.0) & (11.0, 12.0) \\
\hline
$0.01 < z < 0.33$ & 1357 [1647] & 1030 [1231] & 887 [1159] & 663 [899] & 476 [718] & 15 [46] \\
$0.33 < z < 0.86$ & 11596 [15497] & 9007 [11746] & 8602 [11291] & 6516 [9064] & 5947 [7485] & 398 [511] \\
$0.86 < z < 1.00$ & 5504 [6276] & 4785 [4899] & 4789 [4864] & 3862 [4058] & 3037 [3536] & 213 [296] \\
$1.00 < z < 1.60$ & 19735 [30677] & 17267 [23646] & 16931 [23196] & 12922 [19089] & 10647 [16020] & 930 [1413] \\
$1.60 < z < 2.30$ & 14618 [30051] & 15760 [21531] & 16264 [19513] & 11216 [14780] & 9775 [12266] & 910 [1154] \\
$2.30 < z < 2.90$ & 8600 [23970] & 10992 [17226] & 14436 [15560] & 9786 [11479] & 5501 [8064] & 345 [445] \\
$2.90 < z < 4.10$ & 9127 [30446] & 11534 [21764] & 15689 [19401] & 12052 [13707] & 5617 [7723] & 428 \\
$4.10 < z < 4.70$ & 2085 [9080] & 2529 [6445] & 3599 [5645] & 2767 [3766] & 1017 [1590] & 124 \\
$4.70 < z < 5.30$ & 1389 [5536] & 1457 [3917] & 2047 [3404] & 1897 [2215] & 1084 & 89 \\
$5.30 < z < 6.00$ & 446 & 519 & 837 & 822 & 442 & 42 \\
\hline
\end{tabular}
\\
{Notes:} {The values in square brackets are predictions from the stellar mass function (SMF) model from \citet{Weaver_2023}; corrections are only applied when the predicted number counts are higher than counts in the catalog.}
\end{table*}

\begin{table*}[bt]
\centering
\caption{\label{table:N_nonstar} Same as Table \ref{table:N_star}, but for the quiescent selection of galaxies.}
\scriptsize
\centering
\begin{tabular}{c|cccccc}
\hline
\hspace{1cm}$\log_{10}(M/M_\odot) \in$ & (8.5, 8.8) & (8.8, 9.1) & (9.1, 9.5) & (9.5, 10.0) & (10.0, 11.0) & (11.0, 12.0) \\
\hline
$0.01 < z < 0.33$ & 224 & 154 & 163 & 160 [195] & 238 [388] & 18 [73] \\
$0.33 < z < 0.86$ & 976 [1346] & 729 [867] & 761 [954] & 1164 [1371] & 3281 [3670] & 585 [805] \\
$0.86 < z < 1.00$ & 292 [350] & 249 & 325 & 462 [513] & 1534 [1631] & 223 [367] \\
$1.00 < z < 1.60$ & 201 [852] & 241 [517] & 366 [568] & 675 [1086] & 3878 [5113] & 608 [917] \\
$1.60 < z < 2.30$ & 6 & 33 & 53 & 177 [274] & 1860 [2591] & 242 [335] \\
$2.30 < z < 2.90$ & 3 & 6 & 15 & 53 & 676 [859] & 90 [102] \\
$2.90 < z < 4.10$ & 24 & 65 & 116 & 133 [181] & 452 [488] & 96 [101] \\
$4.10 < z < 4.70$ & 11 & 27 & 76 & 86 [133] & 108 & 71 \\
$4.70 < z < 5.30$ & 6 & 18 & 37 & 73 & 357 & 34 \\
$5.30 < z < 6.00$ & 0 & 2 & 14 & 20 & 91 & 26 \\
\hline
\end{tabular}
\\
{Notes:} {The selection is made using a rest-frame two-color $NUV-r$ versus $R-J$ criterion \citep{Ilbert2013, 2013Arnouts}.}
\end{table*}

\end{document}